\documentclass{IEEE-Con-Sys-mag}

\newcommand{\R}{{\mathbb{R}}}
\usepackage{algorithm}
\usepackage{algpseudocode}
\usepackage{hyperref}
\usepackage{url}
\usepackage{mathabx,mathrsfs,stmaryrd}
\newcommand{\real}{{\mathbb{R}}}

\newcommand{\N}{{\mathbb{N}}}

\numberwithin{equation}{section}

 \definecolor{darkgreen}{rgb}{0,0.5,0}

\jvol{XX}
\jnum{XX}
\paper{8}
\jmonth{October}
\jname{IEEE CONTROL SYSTEMS}
\pubyear{2026}

\begin{document}

\begin{titlepage}
\thispagestyle{empty}

\vspace*{2cm}

\begin{center}
    {\Large\bfseries Copyright Notice}
\end{center}

\vspace{1cm}

\noindent
\textcopyright~2026 IEEE. Personal use of this material is permitted.
Permission from IEEE must be obtained for all other uses,
in any current or future media, including
reprinting/republishing this material for advertising
or promotional purposes, creating new collective works,
for resale or redistribution to servers or lists, or
reuse of any copyrighted component of this work in
other works.

\vspace{1cm}

\medskip

\noindent
The published version is available at:
\href{https://doi.org/10.1109/MCS.2026.3715420}{https://doi.org/10.1109/MCS.2026.3715420}

\end{titlepage}

\setcounter{page}{1}

\title{Deep learning from\break a control vantage point
\stitle{Feedforward, residual, transformers, LLMs,\\
and selective state-space models}}

\title{Deep learning from\break  a control vantage point\stitle{Feedforward, residual, transformers, LLMs,\\ and selective state-space models}}

\author{\'A. {Rodr\'iguez Abella}$^\boxempty$, J. P. Silvestre$^\boxdot$, B. Gharesifard$^\boxcircle$, and P. Tabuada$^\boxdot$}
\affil{$^\boxempty$Comillas Pontifical University, $^\boxcircle$Queens University, $^\boxdot$University of California at Los Angeles}

\maketitle


\chapterinitial{D}eep neural networks, in their different shapes and flavors, are having a remarkable impact in almost every area of science and engineering. While they are also influencing the field of control, the purpose of this article is to discuss recent efforts in using ideas from control to better understand and design neural networks. 

The connections between neural networks and control are both old and new. It was already observed in the '80s that backpropagation, the main technique to train neural networks, is closely related to optimal control. 
A different connection emerged more recently, motivated by a class of neural networks termed \emph{residual neural networks}. It was noticed that their update equations are similar to those resulting from the forward Euler discretization of differential equations. This insight motivated the use of differential equation models, and control systems in particular, to describe deep neural networks. Consequently, several  research efforts were initiated to tackle a range of problems in deep learning by leveraging the interplay between control and deep neural networks.


It is from this control vantage point that this  article offers a gentle introduction to deep learning with the objective of making its key ideas more accessible to a control audience while emphasizing how control theory has played, and continues to play, a fundamental role in deep learning. The choice of contents is based on the last author's plenary talk at the 2024 IEEE CDC and it reflects the authors' idiosyncratic view of the area. Any omission of relevant research efforts is simply a consequence of the authors' incomplete view of the rapidly moving interface between control theory and deep learning.  Other articles adopting a control theory perspective include~\cite{zuazua2025machinelearningcontrolfoundations,Han_Li_2022} and complement this article.
)

Notwithstanding the topic of this article, the authors emphatically state that no AI tools were used to generate its contents.

\begin{pullquote}
Notwithstanding the topic of this article, the authors emphatically state that no AI tools were used to generate its contents.
\end{pullquote}

\begin{summary}
\summaryinitial{T}his article offers an overview of recent developments in deep learning from a control vantage point. It reviews several classes of models including feedforward and residual neural networks, transformers, large-language models, and selective state space models while explaining how control theory plays an important role in understanding their training, properties, or architecture design. The article explains how residual neural networks can be interpreted as the discretization of a continuous-time control system thereby enabling the use of control theoretic techniques for its analysis. To illustrate the power of this bridge between control and deep learning, two technical results are discussed: 1) the ability of a residual neural network to approximate arbitrary continuous functions is a non-standard controllability problem that can be solved using geometric nonlinear control and 2) transformers, the key architectural component of most large language-models, can be seen the discretization of a consensus algorithm and thus large language-models exhibit consensus. Another point of contact between control theory and deep learning is the connection between backpropagation, the algorithm used to train neural networks, and optimal control. The article addresses how the older connection between these problems, established in the 80’s using discrete-time models can be reconciled with the more recent approach using continuous-time models. Several open problems and directions for future research are discussed, including selective state-space models, a recent class of models inspired by the models used in control.

\end{summary}

\section{A concise historical account of artificial neural networks}
\label{Sec:Historical}

Despite the recent rise into prominence of neural networks (NNs), their origins date back to the '40s. This section provides a brief overview of the historical context of NNs. Readers not familiar with NNs are encouraged to first read the sidebar ``\nameref{WhatAreNNs1}''.

NNs were originally introduced as simple models of biological neurons and synapses. They were composed of two main components: neurons and the connections between them. An influential early model was introduced by McCulloch and Pitts in 1943~\cite{mcculloch1943logical}, who described neurons as Boolean threshold elements. Building on this line of work, Rosenblatt introduced the perceptron in 1958~\cite{rosenblatt1958perceptron}, usually regarded as one of the first trainable artificial NN architectures. Mathematically, each  neuron can be described by a function transforming the input $z_{\mathrm{in}}\in\R^n$ into the output
\begin{equation}
    z_{out} = \sigma\left( w\, z_{in} + b\right)\in \R,
\end{equation}
where $w\in\mathbb{R}^{1\times n}$ denotes the \emph{weights}, $b\in\mathbb{R}$ denotes the \emph{bias}, and $\sigma:\R\to \R$ is the \emph{activation function} 
\begin{equation}
    \sigma(z)=
    \begin{cases}
        1, & z \geq 0, \\
        0, & z < 0.
    \end{cases}
\end{equation}
The perceptron was a trainable network of such artificial  neurons with supervised learning used to adjust its parameters.

\begin{sidebar}{}
\section{Classification vs regression tasks}
\label{ClassificationVsRegression}

\setcounter{sequation}{0}
\renewcommand{\thesequation}{S\arabic{sequation}}
\setcounter{stable}{0}
\renewcommand{\thestable}{S\arabic{stable}}
\setcounter{sfigure}{0}
\renewcommand{\thesfigure}{S\arabic{sfigure}}

\sdbarinitial{C}lassification and regression are two common machine learning tasks. Both tasks can be seen as the learning of a map $\Phi:X\to Y$.  The goal of \emph{classification} is to separate data into finitely many different classes, hence, $Y$ is a finite set. The classes are said to be linearly separable when $X=\R^m$, for some $m\in \N$, and they can be separated by an affine subspace of $X$.

In a \emph{regression} task, data is to be mapped to real-valued outputs and thus $Y$ is a subset of $\R^n$, for some $n\in \N$.
\end{sidebar}



This network could, in theory, solve any classification task whose classes are linearly separable, see ``\nameref{ClassificationVsRegression}'' sidebar. However, as noted in the book \emph{Perceptrons} (1969)~\cite{minsky1969perceptrons}, it could not solve simple nonlinear problems such as classifying the inputs of the XOR function into the class $\{(0,0), (1,1)\}$, mapped by the XOR function to $0$, and the class $\{(0,1), (1,0)\}$, mapped to $1$. 
This limitation arises from the architectural limitations of single layer perceptron networks, and the lack, at the time, of an effective method  for training networks with multiple layers. The book~\cite{minsky1969perceptrons} is often cited as a contributing factor to the so-called ``AI Winter'', a period where artificial intelligence (AI) research stagnated dramatically. For an overview of how AI relates to NNs and deep learning see the sidebar~``\nameref{AI}''.

\begin{sidebar}{}
\section{AI, machine learning, or deep learning?}
\label{AI}

\setcounter{sequation}{0}
\renewcommand{\thesequation}{S\arabic{sequation}}
\setcounter{stable}{0}
\renewcommand{\thestable}{S\arabic{stable}}
\setcounter{sfigure}{0}
\renewcommand{\thesfigure}{S\arabic{sfigure}}

\sdbarinitial{A}I is the acronym for Artificial Intelligence, a term that encompasses a multitude of techniques aiming at creating machine intelligence. One such technique is machine learning but AI also includes other techniques such as logical and symbolic reasoning. Machine learning focuses on algorithms that learn from data while deep learning is a narrower class of such algorithms utilizing deep neural networks, that is, neural networks with a large number of layers. 
    
\end{sidebar}

The main breakthrough that pushed the field forward was the introduction of backpropagation. This approach introduced differentiable activation functions, such as sigmoids, along with a practical method for training multilayer NNs~\cite{rumelhart1986learning,le1986learning}. Other derivations of backpropagation appeared earlier in different contexts such as \cite{werbos1974beyond}. Although these developments still shared some similarities with biological models, such as the sigmoid function used in the Wilson--Cowan model \cite{wilson1972excitatory}, the connection between artificial NNs and neuroscience gradually weakened. By the late '80s, the two fields had largely diverged, with NNs increasingly driven by empirical results rather than biological realism. Nevertheless, same later models such as spiking NNs~\cite{maass1997networks} retained closer ties to neuroscience, drawing inspiration from classical models including the Hodgkin--Huxley equations \cite{hodgkin1952quantitative}.


As the field progressed, many other NN architectures were introduced. Among the most prominent were recurrent neural networks (RNNs) \cite{elman1990finding}, which were constructed as sequence-to-sequence models, that is, as systems transforming input signals into output signals. RNNs possess, in theory, the ability to retain information over arbitrarily long time horizons, allowing current outputs to be related to past inputs. In practice, this capability is limited by several issues, most notably the vanishing gradient problem (the use of gradients in the training of NNs is discussed in~\emph{\nameref{Sec:OptimalControlBackpropagation}}) \cite{hochreiter2001gradient}. Despite these limitations, RNNs were widely used. 

Among the many different architectures that appeared in the literature, the following three have deep connections to control: residual neural networks, transformers, and selective state-space models. They are described in more detail in the sidebars~``\nameref{WhatAreNNs1}'',~``\nameref{WhatAreNNs1.5}'',~``\nameref{WhatAreNNs2}'', and~``\nameref{WhatAreNNs3}''
\begin{pullquote}
Consequently,~\eqref{ODE} can be regarded as a control system and the problem of training a NN becomes an open-loop control design problem. 
\end{pullquote}

\begin{sidebar}{}
\section{What are feedforward neural networks?}
\label{WhatAreNNs1}

\setcounter{sequation}{0}
\renewcommand{\thesequation}{S\arabic{sequation}}
\setcounter{stable}{0}
\renewcommand{\thestable}{S\arabic{stable}}
\setcounter{sfigure}{0}
\renewcommand{\thesfigure}{S\arabic{sfigure}}

\sdbarinitial{A} neural network (NN) is a family of functions $\Phi_W:\R^{n_1}\to \R^{o}$ parameterized by a set of weights $W$. The function $\Phi_W$ is built by composing simpler functions
\begin{equation}
\label{Composition}
\Phi_W=h\circ g_{\ell-1}\circ g_{\ell-2}\circ \hdots\circ g_1,
\end{equation}
where each $g_k:\R^{n_k}\to \R^{n_{k+1}}$ takes a different form depending on the architecture and $h:\R^{n_\ell} \to \R^o$ is chosen according to the task to be performed.  The length or depth of the NN is denoted by $\ell\in \mathbb{N}$. A NN is said to be a \emph{deep neural network} when $\ell$ is large. The domains and co-domains of the functions $g_i$ are often interpreted as layers, especially when a NN is represented as a graph, see Figure~\ref{Fig:DeepNN}. Under such interpretation, the state $z(k)\in\R^{n_k}$ describes the contents of each of the $n_k$ neurons in layer $k$ and the function $g_k$ transforms the state $z(k)$ of layer $k$ to the state $z(k+1)\in\R^{n_{k+1}}$ of layer $k+1$.
In general, different layers  have different dimensions. For notational simplicity, fix
$n \stackrel{def}{=} \max_k n_k,$
and lift each function $g_k:\R^{n_k}\to \R^{n_{k+1}}$ to the function $\widetilde{g}_k:\R^n\to\R^n$
$$
\widetilde{g}_k(z)=(g_{k,1}(z_1,\hdots,z_{n_k}),\hdots,g_{k,{n_{k+1}}}(z_1,\hdots,z_{n_k}), \underbrace{0,\hdots,0}_{n-n_{k+1}\text{ zeros}})\notag
$$
that has the same domain and codomain. In this article $g_k$ is used to directly denote its lifted version.

\sdbarfig{\includegraphics[width=19.0pc]{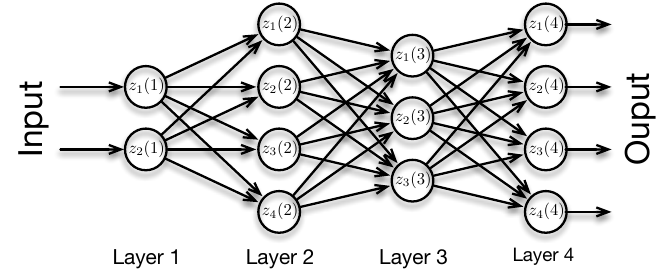}}{Neural network depicted as a graph. The vertically stacked circles represent neurons and form a layer. The arrows from layer $k$ to layer $k+1$ represent the different entries of the weight matrix $W(k)$.\label{Fig:DeepNN}}

\subsection{Feedforward Networks}
Feedforward networks are among the earliest NN architectures and still widely used. In a feedforward NN, data moves only in one direction: it enters at the first or input layer as $z(1)$, is successively transformed by the sequence of functions $g_k$ producing $z(k+1)$, and results in the output $h(z(\ell))$. In feedforward NNs the maps $g_k$ take the form
\begin{equation}
\label{StateUpdateSideBar}
g_k(z)=\Sigma(W(k)z),
\end{equation}
where the function $\Sigma:\R^n\to\R^n$ operates element-wise, that is
$$\Sigma(v)=(\sigma(v_1),\sigma(v_2),\hdots,\sigma(v_n))$$
with $\sigma:\R\to\R$ an activation function. The argument of $\Sigma$ is the linear function of $z$ defined by the weights $W(k)\in \R^{n\times n}$. In the literature, the argument of $\Sigma$ is often an affine function $W(k)z+b(k)$. To simplify the presentation, a linear representation is used; it can be obtained by writing $W(k)z+b(k)$ in the linear form 
$$\begin{bmatrix}
W(k) & b(k)\\
0 & 1\end{bmatrix}
\begin{bmatrix}
z\\1
\end{bmatrix}.$$
In this article $W$ and $z$ are used to directly denote their lifted variants.




Classical feedforward NNs used the logistic sigmoid \mbox{$\sigma(v_i)=(1+e^{-v_i})^{-1}$} or the hyperbolic tangent \mbox{$\sigma(v_i)=\tanh(v_i)$} as activation functions. While smooth, these functions can saturate and attenuate gradients thus making training more challenging. To facilitate deeper architectures, modern networks predominantly employ the Rectified Linear Unit (ReLU), $\sigma(v_i)=\max\{0,v_i\}$ \cite{nair2010rectified}, and its variants. More recently, smooth gating-style activations like the Gaussian Error Linear Unit (GELU) \cite{hendrycks2016gaussian}, defined as $\sigma(v_i)= \frac{v_i}{2}\left(1+\operatorname{e}\!\left(\frac{v_i}{\sqrt{2}}\right)\right)$,  $\operatorname{e}(v_i) = \frac{2}{\sqrt{\pi}}\int_{0}^{v_i} e^{-t^{2}}\,dt$,  have become standard in transformer-based models.

\subsection{Task Dependence}
The choice of the function $h$ in~\eqref{Composition} must be made according to the task being solved. Two common cases are:

\emph{Regression:} In a regression problem the output of a NN is an element of $\R^o$ and it is common to choose $h$ to be affine
\begin{equation}
h(z(\ell))= A z(\ell) + c,    
\end{equation}
where $A\in\mathbb{R}^{o\times n}$ and $c\in\mathbb{R}^o$ are weights learned from data.

\emph{Classification:} In a classification problem, the output is the class to which the input belongs. Rather than directly producing the class, NNs produce a vector that is interpreted as the probability of the input belonging to each class. Hence, if there are $o\in \mathbb{N}$ classes, the output is a vector in
$$
\Delta_o \stackrel{def}{=} \left\{p\in\R^o:\ p_i\ge 0,\ \sum_{i=1}^o p_i=1\right\}.
$$
The output map $h$ in~\eqref{Composition} is the composition of two maps. The first is an affine map acting on $z(\ell)$ and producing \emph{logits} $q\in \R^o$
$$q=Az(\ell)+c,$$
and the second converts the logits into an element $p\in \Delta_o$ using the softmax function
\begin{equation}
p_i = \mathrm{softmax}_i(q) \stackrel{def}{=} \frac{e^{q_i}}{\sum_{j=1}^o e^{q_j}},\qquad i\in\{1,\dots,o\}.
\end{equation}
Each $p_i$ is then interpreted as the probability of the given input $z(1)$ belonging to class $i$. 

A more detailed introduction to this class of networks can be found in ~\cite[Chapter~6]{goodfellow2016deep}.

\end{sidebar}

\begin{sidebar}{}
\section{What are residual neural networks?}
\label{WhatAreNNs1.5}

\setcounter{sequation}{0}
\renewcommand{\thesequation}{S\arabic{sequation}}
\setcounter{stable}{0}
\renewcommand{\thestable}{S\arabic{stable}}
\setcounter{sfigure}{0}
\renewcommand{\thesfigure}{S\arabic{sfigure}}


\sdbarinitial{I}n recent years, the surge in computational power and large-scale datasets has enabled the training of substantially deeper networks, beginning with milestones such as AlexNet~\cite{krizhevsky2012imagenet}, which established convolutional neural networks as the standard for visual recognition. However, as depth increases, the training of such networks becomes notoriously difficult. Deep residual networks (ResNets)~\cite{he2016deep} addressed this by introducing residual (skip) connections, which allow the gradient to bypass intermediate layers, see Figure~\ref{Fig:Skip}.

Formally, the functions $g_k$ (see~\eqref{Composition}) in a ResNet are of the form
\begin{equation}
g_k(z)= z + W_2(k)\Sigma(W_1(k) z),
\label{eq:residual_connection}
\end{equation}
where $W_1(k), W_2(k)\in \R^{n\times n}$ are weight matrices. It has been empirically shown that this architectural change, with respect to feedforward NNs, effectively smoothens the training loss landscape and mitigates the vanishing gradient problem~\cite{li2018visualizing} (the use of gradients in the training of NNs is discussed in~\emph{\nameref{Sec:OptimalControlBackpropagation}}) thereby enabling the reliable training of ResNets with hundreds of layers. 

\sdbarfig{\includegraphics[width=19.0pc]{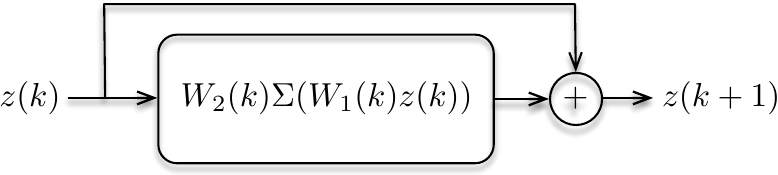}}{Graphical description of a skip connection in a ResNet\label{Fig:Skip}.}

\end{sidebar}

\begin{sidebar}{}
\section{What are transformers?}
\label{WhatAreNNs2}

\setcounter{sequation}{0}
\renewcommand{\thesequation}{S\arabic{sequation}}
\setcounter{stable}{0}
\renewcommand{\thestable}{S\arabic{stable}}
\setcounter{sfigure}{0}
\renewcommand{\thesfigure}{S\arabic{sfigure}}


\sdbarinitial{A} common requirement in modern applications of NNs is to approximate functions mapping signals to signals, also known as sequence-to-sequence mappings. One example is the translation of text from one language to another. In such settings, the input is a sequence
\begin{equation}
\label{Sequence1}
 (s_1,\dots,s_m),\qquad s_i\in\mathbb{R}^{n}
\end{equation}
and the output can be a  sequence or a classification label. It has been empirically shown that the transformer architecture~\cite{vaswani2017attention} outperforms feedforward NNs in learning sequence-to-sequence maps. Transformers handle sequences  by building intermediate token representations and repeatedly mixing information across different elements of the sequence using an \emph{attention} mechanism. Attention mechanisms were popularized in neural machine translation \cite{bahdanau2014neural} and later instantiated as \emph{self-attention} in transformers \cite{vaswani2017attention}. This sidebar provides a brief introduction to these concepts.

The transformer architecture can still be described by a map similar to~\eqref{Composition}
\begin{equation}
\label{TransformerEq}
\Phi_W=h\circ g_{\ell-1}\circ g_{\ell-2}\circ \hdots\circ g_1\circ f.
\end{equation}
The function $f:\R^{n\times m}\to \R^{d\times m}$ is called the embedding map and sends the matrix representation of the sequence~\eqref{Sequence1}
\begin{equation}\label{Sequence2}
S=\begin{bmatrix}s_1\vert s_2\vert\hdots\vert s_m\end{bmatrix}\in\R^{n\times m}
\end{equation}
into a token matrix in $\R^{d\times m}$. The maps $g_k:\R^{d\times m}\to\R^{d\times m}$ describe the transformer blocks, and the map $h$  is the output map. The domain of $h$ is $\R^{d\times m}$ and its co-domain depends on the task.

\textbf{Embedding and positional information:} The embedding map $f$ is typically of the form
$$f(S)=\begin{bmatrix}E(s_1,1)\vert E(s_2,2)\vert\hdots\vert E_m(s_m,m)\end{bmatrix}$$
for a function $E:\R^n\times \mathbb{N}\to \R^d$ whose form depends on the specific application and whose parameters are learned from  data. For the sake of brevity, we also call each element
\begin{equation}
z_i(1) = E(s_i ,i)
\end{equation}
a token although the term token is often reserved for the elements $s_i$. Collecting tokens column-wise leads to the token matrix representation
\begin{equation}
    Z(1) \stackrel{def}{=} \begin{bmatrix}
    z_1(1)\vert & \dots&\vert  z_m(1)
\end{bmatrix}\in \R^{d\times m}
\end{equation}
that will be used in the next sections.

\textbf{Self-attention and the transformer block}

At each layer $k\in\{1,\dots,\ell-1\}$, the transformer updates $Z(k)\in\mathbb{R}^{d\times m}$ using multi-head self-attention followed by a token-wise feedforward NN layer, each combined with a residual connection and a normalization function \cite{vaswani2017attention}.

Denote by $a\in\N$ the total number of heads in a transformer. A head is simply a parallel path in the architecture along which an attention mechanism is implemented.  The output of all the heads is then combined into a single output. Define the key $K_j(k)\in\R^{e\times m}$, query $Q_j(k)\in\R^{e\times m}$, and value $V_j(k)\in\R^{e'\times m}$ matrices for the $j$-th head as
\begin{align}
    Q_j(k) & = W_{Q_j}(k) Z(k),\notag\\
    K_j(k) & = W_{K_j}(k) Z(k),\notag\\
    V_j(k) & = W_{V_j}(k) Z(k),\notag
\end{align}
where $W_{Q_j}(k),W_{K_j}(k)\in\mathbb{R}^{e\times d}$ and $W_{V_j}(k)\in\mathbb{R}^{e'\times d}$ are weight matrices. The output of the attention mechanism is given by summing over all the attention heads
\begin{equation}
\label{Attention}
A(Z(k)) = \sum_{j=1}^a W_{o_j}V_j(k)\mathrm{softmax}\!\left(\frac{K_j^\top(k) Q_j(k)}{\sqrt{e}}\right),
\end{equation}
where softmax is applied column-wise and $W_{o_j}\in\R^{d\times e'}$ is another weight matrix. The matrix $K_j^\top(k) Q_j(k)$ contains the inner products between the columns of $Z$ when using the inner product defined by the weight matrix $W_{K_j}^\top(k) W_{Q_j}(K)$. In particular, the entry $ (i_1,i_2) $ of $K_j^\top(k) Q_j(k)$ measures how much token $ i_1 $ is aligned with token $i_2$. The division by $\sqrt{e}$ prevents the inner products in $K_j^\top(k) Q_j(k)$ from growing with the dimension. The standard transformer block update, that is, the map $g_k$ in~\eqref{TransformerEq}, can then be written as the composition of two maps:
\begin{align}
\widetilde Z(k) &= N\!\big(Z(k) + A(Z(k))\big),\label{eq:attn_res_norm}\\
Z(k+1) &= N\!\big(\widetilde Z(k) + R_k(\widetilde Z(k))\big),\label{eq:ffn_res_norm}
\end{align}
where $N:\R^{d\times m}\to\R^{d\times m}$ is a normalization map applied token-wise (column-wise) and $R_k(Z(k))=(r_k(z_1),\dots,r_k(z_m))$, with $r_k$ defined by
$$r_k(z_i)=  W_2(k)\Sigma(W_1(k)z_i),$$
as depicted in Figure~\ref{Fig:Attention}.

In the original transformer architecture, there are two types of transformer blocks:  encoders and decoders. In the encoder blocks, attention is typically computed over all token positions $1,\dots,m$. In the decoder blocks, attention is made \emph{causal}, that is,~\eqref{Attention} is modified so that attention for token $i$ only depends on the tokens that precede it, that is, tokens $1$ through $i-1$ ~\cite{vaswani2017attention}.

\textbf{Output and text generation:} The transformer output is task-dependent. For classification, the output map $h$ is typically the composition of two maps, similarly to what is done in feedforward NNs, see sidebar~``\nameref{WhatAreNNs1.5}''. The first is an affine map acting on a designated token  (often the final token or a special classification token) to obtain logits. The second map is the entry-wise softmax to produce a distribution over classes. 

For text generation, the output map is similarly defined but the class set is the vocabulary of the language being used. The output map sends a token matrix to logits over all the vocabulary words, and the softmax converts these logits into a probability distribution for the next token. The next token is then selected by a decoding rule, passed through the embedding map $f$ and appended to the end of the token matrix $Z(1)$. The process is then restarted by processing the enlarged token matrix starting with the map $g_1$. This text generation process ends when a special \emph{end of sentence} or \emph{end of text} token is generated.

In practice, decoding often modifies the softmax distribution before sampling. Two common  rules are:
(i) \emph{top-$k$} sampling, which restricts sampling to the $k$ most probable tokens; and
(ii) \emph{top-$p$} sampling, which restricts sampling to the smallest set of tokens whose cumulative probability is at least $p$, see \cite{holtzman2019curious}.

Readers seeking additional background on large language models and transformers may consult \cite[Chapter~7-8]{jm3}.

\sdbarfig{\includegraphics[width=22.0pc]{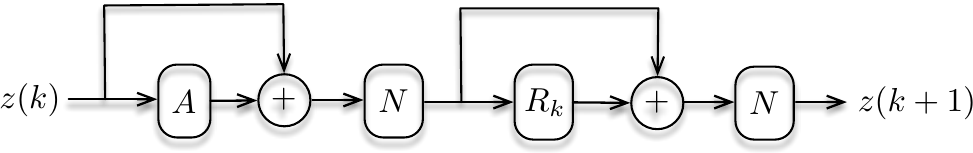}}{Diagrammatic representation of a transformer block architecture described by~\eqref{eq:attn_res_norm} and~\eqref{eq:ffn_res_norm}.\label{Fig:Attention}} 
\end{sidebar}

\begin{sidebar}{}
\section{What are recurrent neural networks and selective state-space models?}
\label{WhatAreNNs3}

\setcounter{sequation}{0}
\renewcommand{\thesequation}{S\arabic{sequation}}
\setcounter{stable}{0}
\renewcommand{\thestable}{S\arabic{stable}}
\setcounter{sfigure}{0}
\renewcommand{\thesfigure}{S\arabic{sfigure}}

\subsection{Recurrent Neural Networks}

An early architecture for sequence-to-sequence modeling is the recurrent neural network (RNN) \cite{elman1990finding}. All the NN architectures so far described were formulated as composition of maps that update the state from one layer to the next. These maps were \emph{memoryless} whereas RNNs employ \emph{memoryfull} maps defined by a nonlinear control system. The memory of such maps is denoted by $c(k,i)\in\mathbb R^d$ with $k$ denoting the dependence on the layer and $i$ denoting the dependence on the position (in the input sequence) of the token being processed. 
As illustrated in Figure~\ref{Fig:RNN}, a RNN can be can interpreted as a two-dimensional architecture in which information moves horizontally, from layer $1$ to layer $\ell$, and also vertically, from position $1$ to position $m$.   The following update equations explain how the memory and state are updated
\begin{equation}
\begin{aligned}
c(k,i) &= \Sigma\!\left(A(k)c(k,i-1) + B(k)z(k,i)\right),\\
z(k+1,i) &= \Sigma\!\left(C(k)c(k,i)\right),\\
c(k,0)&= 0,
\end{aligned}
\label{eq:rnn}
\end{equation}
for matrices $A(k)\in\R^{d\times d}$,~$B(k)\in\R^{d\times n}$ and~$C(k)\in\R^{n\times d}$.  In practice, RNNs can be difficult to train  due to the well known vanishing/exploding gradient problem. Some variants such as Long Short-Term Memory (LSTM) networks \cite{hochreiter1997long} can mitigate this problem, but the training process remains inherently sequential in $i$ thus limiting parallelization. 

A good introduction to  this class of models can be found in~\cite[Chapter~10]{goodfellow2016deep}.

\subsection{Selective State-Space Models}

More recently, \emph{selective state-space models} (SSMs) have emerged as an alternative approach to transformers \cite{gu2021efficiently,smith2022simplified}. Among all the different models, which are explained in depth in \cite{alonso2025state}, the most prominent one is the Mamba model~\cite{gu2024mamba} whose defining feature is the use of input dependent weight matrices.

\sdbarfig{\includegraphics[width=19.0pc]{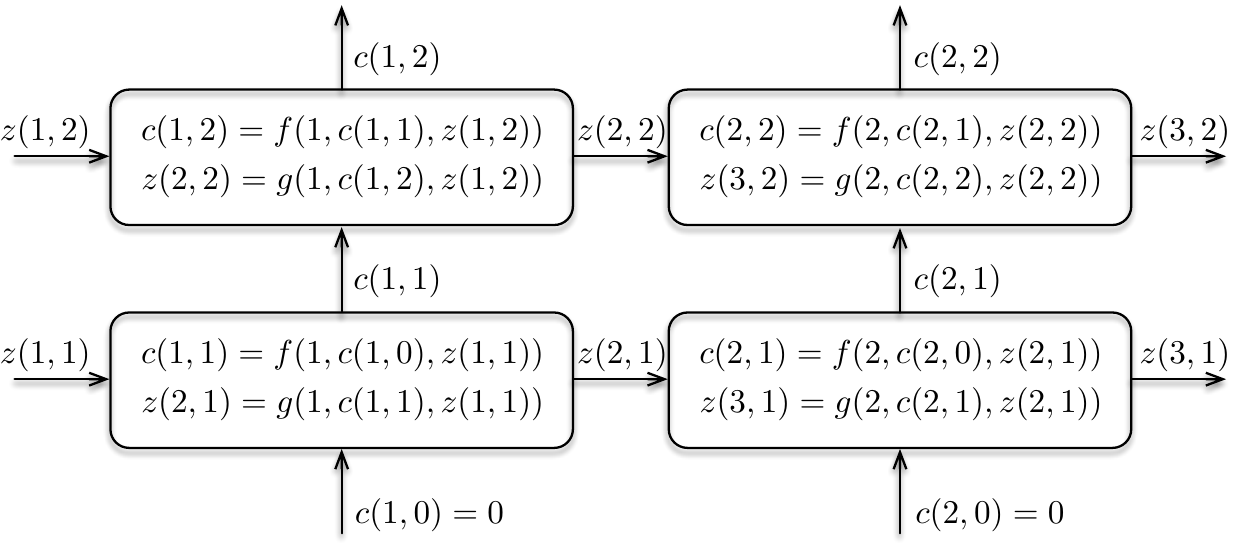}}{Graphical description of the information flow in an SSM. The update equations~\eqref{eq:ssm_selective_compact} were simplified as follows $f(k,c(k,i-1),z(k,i))=A\!\left(k,z(k,i)\right)\,c(k,i-1) + B\!\left(k,z(k,i)\right)\,z(k,i)$ and $g(k,c(k,i),z(k,i))=C\!\left(k,z(k,i)\right)\,c(k,i)$. Note that $c(1,0)$ and $c(2,0)$ are only used for the purpose of initialization since position $i=0$ in the input string does not exist. \label{Fig:RNN}}
Its update equations are
\begin{equation}
\begin{aligned}
c(k,i) &= A\!\left(k,z(k,i)\right)\,c(k,i-1) + B\!\left(k,z(k,i)\right)\,z(k,i),\\
z(k+1,i) &= C\!\left(k,z(k,i)\right)\,c(k,i),\\
c(k,0) &=0,
\end{aligned}
\label{eq:ssm_selective_compact}
\end{equation}
where $A:\{1,\hdots,\ell\}\times \R^n\to \R^{d\times d}$, $B:\{1,\hdots,\ell\}\times \R^n\to \R^{d\times n}$, and $C:\{1,\hdots,\ell\}\times \R^n\to \R^{n\times d}$ are nonlinear maps whose parameters are learned from data. The dependence on the layer state (which is the input for layer $1$) allows the model to adapt the weights to the current input thereby changing how the state propagates. Consequently, more relevant inputs can have more impact on the state whereas less relevant input information can be discarded \cite{gu2024mamba}. 

Empirically, selective SSMs such as the Mamba can have a comparable performance to transformers \cite{gu2024mamba}. Moreover, under certain choices for the matrices $A(k,z(k,i))$, $B(k,z(k,i))$, and $C(k,z(k,i))$ the computational complexity of training these models is lower than transformers. Finally, it is possible to interleave SSM blocks with attention  blocks to form hybrid architectures such as the Jamba model \cite{lenz2025jamba}.

\end{sidebar}

\section{Differential equations, control systems, and neural networks}\label{DiffEqs}

As discussed in \emph{\nameref{Sec:Historical}}, differential equations were already used in the '50s to model \emph{biological} NNs. Surprisingly, it was recently observed that differential equations can also be used to describe \emph{artificial} NNs. This recent connection was motivated by an architecture called residual networks (ResNets), see sidebar~``\nameref{WhatAreNNs1.5}'' for a more detailed description of ResNets. 

What distinguishes ResNets from other architectures is how the state at layer $k\in\{1,\hdots,\ell-1\}$, denoted by $z(k)\in \R^n$, gives rise to the state at layer $k+1$
\begin{equation}
\label{ResNetStateUpdate}
z(k+1)=z(k)+w(k)\Sigma(W(k)z(k)).
\end{equation}
The function $\Sigma:\R^n\to\R^n$ describes the action of an activation function $\sigma:\R\to\R$ on each coordinate
$$\Sigma(v)=(\sigma(v_1),\sigma(v_2),\hdots,\sigma(v_n)).$$
The argument of $\Sigma$ in~\eqref{ResNetStateUpdate} is the linear function of $z(k)$ defined by the weights $W(k)\in \R^{n\times n}$. Note also the presence of the scalar weight $w(k)\in\R$. Although  the original formulation in~\cite{he2016deep} allows for $w$ to be a matrix, it is pedagogical to focus on the scalar case.

It was observed in~\cite{E2017,HaRu2017,LuZLD18,10.5555/3327757.3327764} that~\eqref{ResNetStateUpdate} has a striking similarity with the numerical solution of the differential equation
\begin{equation}
\label{ODE}
\dot{z}_c(t)=\Sigma(W_c(t)z_c(t))
\end{equation}
given by the forward Euler method
$$z(t_k+\tau)=z(t_k)+\tau\Sigma(W_c(t_k)z(t_k)),$$
where $\tau\in \R^+$ is the step size and $t_k=k\tau$. Hence, if the weights $w(k)$ are small, it is conceivable that ResNets can be regarded as the discretization of~\eqref{ODE}. Before elaborating further on the possible equivalence between ResNets and differential equations, it matters to delve into the control perspective afforded by this connection. In~\eqref{ODE}, $W_c$ is a map $W_c: [0,1]\to \R^{n\times n}$ transforming time $t\in[0,1]$ into a matrix of weights. For convenience,  the time domain was normalized to $[0,1]$. The map $W_c$ is to be learned from data, that is, it is to be designed. This suggests that $W_c$ can be interpreted as a control input. Consequently,~\eqref{ODE} can be regarded as a control system and the problem of training a NN becomes an open-loop control design problem. 
This perspective brings into clear view how control theory can play a fundamental role in deep learning. Examples of how this perspective brought solutions to new and existing problems include an alternative connection between backpropagation and control (an earlier connection was identified in the '80s), establishing deep neural networks as universal function approximators, and explaining the attention mechanism in transformers through consensus. 

Since the interplay between control systems and ResNets is the unifying theme of this article, a detailed  discussion of when  ResNets can be seen as the discretization of differential equations, or a differential equation can be seen as the continuous limit of a ResNet, is in order. Although outside the scope of this article, there is also a parallel line of work relating NNs and partial differential equations~\cite{li2020fourier,kovachki2023neural,krstic2024machine}.

\subsection{Fixed inputs}
Consider first the case where the control input \mbox{$W_c:[0,1]\to \R^{n\times n}$} is known and fixed where we normalized time to the interval $[0,1]$ for simplicity. This represents situations where analysis or design was conducted using the model~\eqref{ODE} and the resulting conclusions are to be carried over, or implemented, on a ResNet. One example is the use of~\eqref{ODE} to establish existence of inputs guaranteeing universal approximation, as discussed in~\emph{\nameref{Sec:UniversalApproximation}}, and then seeking to determine if ResNets have the same ability. Using the Euclidean norm, denoted by $\Vert\cdot\Vert$, questions of this type can be distilled to:

\emph{
Given a desired error tolerance $\varepsilon \in \R^+$, is there a number of layers $\ell\ge 2$ and a choice of weights $w$ so that the mismatch between the solution $z_c:[0,1]\to \R^n$ of $\dot{z}_c(t)=\Sigma(W_c(t)z_c(t))$ and the solution $z:\{1,2,\hdots,\ell\}\to \R^n$ of the recursion
\begin{align}
\label{Recursion}
z(1)&=z_c(0),  \notag\\
z(k+1)&=z(k)+w(k)\Sigma\left(W_c\left(\frac{k-1}{\ell-1}\right) z(k)\right),
\end{align}
satisfies
\begin{equation}
\label{EpsilonClose}
\left\Vert z_c\left(\frac{k-1}{\ell-1}\right)-z(k)\right\Vert\le \varepsilon 
\end{equation}
for every $k\in\{1,\hdots,\ell\}$?
}

 Note how the control input $W_c:[0,1]\to \R^{n\times n}$ directly defines the weights  $W:\{1,2,\hdots,\ell\}\to \R^{n\times n}$ of the ResNet
$$W(k)=W_c\left(\frac{k-1}{\ell-1}\right)$$
and the only degrees of freedom are the number of layers $\ell$ and the weights $w$ that play the role of step size. This is a classical problem in the numerical analysis of differential equations and the answer is affirmative under suitable technical conditions, see, for example~\cite{NumAnal}. A particular choice of such conditions is continuous differentiability of $\Sigma$, boundedness of $W_c$, and the assumption that~\eqref{EpsilonClose} only needs to hold for solutions with $z_c(0)$ belonging to a compact set. Note that implicit in the problem formulation is that solutions of~\eqref{ODE} exist in the time domain $[0,1]$. In general, lower values of $\varepsilon$ will require larger $\ell$, that is, deeper ResNets. Since $\varepsilon$ in~\eqref{EpsilonClose} can be rendered as small as desired,  the differential equation  model~\eqref{ODE} can be seen as the limit of a ResNet when $\ell\to \infty$ for a suitable choice of weights $w$ such as $w(k)=1/\ell$.

\subsection{Learned inputs}
Whereas the case of known and fixed control inputs can be resolved using numerical analysis, there are situations where the control inputs are not known beforehand. One such example occurs when the model~\eqref{ODE} is used to design an algorithm that learns $W_c:[0,1]\to \R^{n\times n}$ from data. Denote such algorithm by a map
$$\phi_c:(\R^n\times \R^n)^s\to (\R^{n\times n})^{[0,1]}$$ 
sending a data set  $\mathcal{D}\in (\R^n\times \R^n)^s$ into $W_c=\phi_c(\mathcal{D})$. The notation $(\R^{n\times n})^{[0,1]}$ denotes the set of all maps from $[0,1]$ to $\R^{n\times n}$ and a data set is a collection of pairs $\mathcal{D}=((x_1,y_1),(x_2,y_2),\hdots,(x_s,y_s))$ where $y_i$ is the evaluation of some unknown function on the data point $x_i$. When using supervised learning, the objective of training a NN is to learn the values of the weights so that the NN approximates the unknown function. Training of NNs is described in more detail in~\emph{\nameref{Sec:OptimalControlBackpropagation}}. Typically, $\phi_c$ models the construction of $W_c$ through the minimization of a loss functional based on the training data
\begin{equation}
\label{ContinuousLoss}
W_c=\mathrm{argmin}_{\widehat{W}_c}\int_0^1 L_c\left(\mathcal{D},\widehat{W}_c(t)\right) dt.
\end{equation}
In order to implement such algorithm,~\eqref{ODE} is discretized to obtain a ResNet and $\phi_c$ is discretized to obtain a family of maps
$$\phi_\ell:(\R^n\times \R^n)^s\to (\R^{n\times n})^{\{1,2,\hdots,\ell\}}$$ 
parameterized by the length $\ell$ of the network. When $\phi_c$ is defined through~\eqref{ContinuousLoss}, $\phi_\ell$ would also be defined by the minimization of a suitably discretized loss
\begin{equation}
\label{DiscreteLoss}
W=\mathrm{argmin}_{\widehat{W}}\sum_{k=1}^{\ell-1} L_\ell\left(\mathcal{D},\widehat{W}\left(k\right)\right).
\end{equation}

It is natural to question if the mismatch between these models decreases as $\ell$ increases. In particular, to question if the weights obtained by minimizing~\eqref{DiscreteLoss} converge to the weights minimizing~\eqref{ContinuousLoss} as $\ell\to\infty$. Note, however, that in this case the control input $W$ for each ResNet does not come from the discretization of a continuous control input $W_c$, which was the setting discussed in the previous section.

  Earlier solutions to this problem were investigated in the literature on numerical methods for optimal control problems, see for example~\cite{2e4cbc61-c1d8-33ae-b1ab-dc7f49a7534b,RungeKuttaOptimalControl,doi:10.1137/140999621}. A more detailed discussion of how optimal control relates to the learning of the weights $W$, appears in~\emph{\nameref{Sec:OptimalControlBackpropagation}}. For now, it suffices to observe that the minimization problems~\eqref{ContinuousLoss} and~\eqref{DiscreteLoss} can be interpreted as optimal control problems. The results in~\cite{doi:10.1137/140999621} establish that the weights $W$ obtained by solving~\eqref{DiscreteLoss} and the evolution of the state of a ResNet when using such weights converge to the control input obtained by solving~\eqref{ContinuousLoss} and to the solution of~\eqref{ODE} when using such control input. These convergence results are predicated on the assumption that the control input obtained by solving~\eqref{ContinuousLoss} is Lipschitz continuous (or on other assumptions implying Lipschitz continuity). The recent paper~\cite{GammaConv} revisits this problem using more recent mathematical tools and forgoes the Lipschitz continuity assumption. Instead, it describes how to regularize the continuous~\eqref{ContinuousLoss} and the discrete loss~\eqref{DiscreteLoss}  to ensure convergence. Intuitively, the regularizer penalizes  weights that change too fast across layers which can be interpreted as penalizing the discretization of the derivative of the continuous inputs. Another interesting point of departure in the paper~\cite{GammaConv} is that it does not rely on Pontryagin's optimality conditions and, instead, it employs the notion of $\Gamma$-convergence  enabling the direct comparison between the  minimizers of the regularized losses~\eqref{ContinuousLoss} and~\eqref{DiscreteLoss} as $\ell\to\infty$.

\subsection{Novel architectures based on different discretization schemes}
The link between ResNets and the differential equation~\eqref{ODE} led several researchers to investigate if discretization schemes beyond the forward Euler method could be used to produce new NN architectures with better properties. Some of the new architectures produced by this process include PolyNets~\cite{ZhLiChLi2016}, based on a polynomial approximation of the backward Euler method, 
a Hamiltonian inspired architecture~\cite{HaRu2017}, based on the semi-implicit Euler method that is known to be symplectic, the momentum ResNet~\cite{SaAbBlPe2021}, based on the Verlet integration scheme, and several others~\cite{LuZLD18,GoReUrGr2017, li2018maximum,ChMeHaRuBeHo2018} based on generalizations of the differential model \eqref{ODE} that include second order derivatives.

Rather than starting with a differential equation and discretizing it to arrive at a NN architecture, the differential equation can be directly taken as the architecture. This is the key idea behind Neural ODEs~\cite{10.5555/3327757.3327764}, in which the NN directly parameterizes the right hand side of a differential equation. The evaluation of a Neural ODE on a given input consists in numerically solving the corresponding differential equation using the input as initial condition. The objective of training is to find the differential equation whose solutions best match the training data. Similarly to evaluation, the training of Neural ODEs requires the use of a numerical solver for differential equations. Neural ODEs are especially useful when working with data that is known to be produced by a differential equation but the use of numerical solvers makes them less scalable than existing alternatives.

\section{Optimal control and backpropagation}
\label{Sec:OptimalControlBackpropagation}
\subsection{The classical connection}
\label{SSec:ClassicalConnection}
The classical connection between control and backpropagation is known since the '80s~\cite{rumelhart1986learning,lecun1988theoretical}. To elucidate it, recall the setting in~``\nameref{WhatAreNNs1}'' for a NN with $\ell\in \mathbb{N}$ layers, all of which have $n$ neurons. The state at layer $k+1\in\{2,\hdots,\ell\}$ is computed from the state at layer $k$ using 
\begin{equation}
\label{StateUpdate}
z(k+1)=\Sigma(W(k)z(k))
\end{equation}
and a NN can be seen as a function $\Phi_W:\R^n\to \R^n$ mapping $z(1)$ to $\Phi_W(z(1))=z(\ell)$, where $z(\ell)$ is recursively computed using~\eqref{StateUpdate}.  NNs are trained, that is, their weights $W$ are computed so that $\Phi_W$ approximates an unknown function for which  \emph{samples} are available. A collection of $s$ samples $((x_1,y_1),(x_2,y_2),\hdots,(x_s,y_s))$ consists of pairs $(x_i,y_i)\in\R^{n}\times\R^n$ where $y_i$ represents the evaluation of the unknown function at $x_i$. Supervised training of a neural network entails  learning the weights $W$ so that $\Phi_W$ maps $x_i$ to $y_i$. 

The learning objective can be formalized as the minimization of the loss functional
$$
\sum_{i=1}^s L(y_i,\Phi_W(x_i))$$
subject to the constraints described by the equalities~\eqref{StateUpdate} and where $L:\R^n\times\R^n\to \R_0^+$ is a differentiable function measuring the discrepancy between $y_i$ and $\Phi_W(x_i)$, such as $\Vert y_i-\Phi_W(x_i)\Vert^2$. 
Incorporating the constraints~\eqref{StateUpdate} into the loss leads to the expanded loss functional
\begin{align}
\mathcal{J}(\lambda,z,W)&=\sum_{i=1}^s L(y_i,\Phi(x_i))\notag\\
&+\sum_{i=1}^s\sum_{k=2}^\ell \lambda^T_i(k)\left[z_i(k)-\Sigma(W(k-1)z_i(k-1))\right]\notag
\end{align}
with $\lambda_i(k)\in\R^n$ being Lagrange multipliers (adjoint variables). This is now an unconstrained minimization problem for which necessary conditions for optimality can be written.

The first necessary condition for optimality arises by differentiating the cost with respect to the Lagrange multipliers
\begin{equation*}
\frac{\partial \mathcal{J}}{\partial \lambda_i(k)}
 = 0
\quad\implies\quad
z_i(k) = \Sigma(W(k-1)z_i(k-1)),
\end{equation*}
and recovers~\eqref{StateUpdate} as expected. The second arises by differentiating the cost with respect to the state
\begin{equation}
\label{2ndNecCond}
\frac{\partial L}{\partial z_i(k)}
 = 0.
\end{equation}
In particular, at the final layer $k=\ell$, it provides
\begin{equation*}
\frac{\partial \mathcal{J}}{\partial z_i(\ell)}
= \frac{\partial L}{\partial z_i(\ell)} + \lambda_i(\ell) = 0,
\end{equation*}
defining the value of $\lambda_i$ at the last layer 
\begin{equation}
\label{FinalCondition}
\quad
\lambda_i(\ell) =-\frac{\partial L}{\partial z_i(\ell)}.
\end{equation}
To obtain $\lambda_i$ at the intermediate layers, 
condition~\eqref{2ndNecCond} is used for $k=2,\hdots,\ell-1$
\begin{align}
\frac{\partial \mathcal{J}}{\partial z_i(k)}&=\lambda_i(k)-\frac{\partial \Sigma}{\partial v}W(k)\lambda_i(k+1)=0,\notag
\end{align}
giving rise to the recursion
\begin{equation}
\label{AdjointEq}
\lambda_i(k)=\frac{\partial \Sigma}{\partial v}W(k)\lambda_i(k+1).
\end{equation}
Note that it is possible that $ \lambda_i(k+1) $ is in the kernel of $W(k)$, leading to vanishing of the adjoint along the NN. In what follows, as it is commonly done in the literature, the weight matrix is assumed not to be rank deficient.

To compute the last necessary condition, note that the derivative of $\Sigma(v)$ with respect to $v$ is the diagonal matrix whose entry in row and column $j$ is $\frac{d\sigma}{d v_{j}}$, $ j \in \{1,\ldots, n\}.$ 
Next, define
\begin{equation}\label{eq:fwd_v}
v_i(k+1) \stackrel{def}{=} W(k) z_i(k)
\end{equation}
for $ i \in\{1,\ldots, s\}$ and note that the $j$-th component of $ v_i (k+1)$ is
\[
v_{i,j}(k+1) = \sum_{m=1}^n W(k)_{jm} z_{i,m}(k).
\]
Consequently
\begin{equation}\label{eq:dervij}
\frac{\partial v_{i,j}(k+1)}{\partial W(k)_{ab}} = \delta_{ja} z_{i,b}(k),    
\end{equation}
$a,b \in \{1,\ldots, n\} $, and where $ \delta_{ja} $ is Kronecker's delta. Using this equality and the definition of $ \mathcal{J} $, the derivative of $ \mathcal{J} $ with respect to the entry in row $a$ and column $b$ of $W$ can be computed 
\begin{align}
    \frac{\partial \mathcal{J}}{\partial W(k)_{ab}}
    &= 
    \frac{\partial}{\partial W(k)_{ab}} \Big(\sum_{i=1}^s-\lambda_i(k+1)^T \Sigma(v_i(k+1))\Big)\nonumber\\
    &= -\sum_{i=1}^s\sum_{j=1}^n \lambda_{i,j}(k+1) \frac{d\sigma}{d v_{i,j}(k+1)} \frac{\partial v_{i,j}(k+1)}{\partial W(k)_{ab}}\nonumber\\
    &= -\sum_{i=1}^s \lambda_{i,a}(k+1) \frac{d\sigma}{d v_{i,a}(k+1)} z_{i,b}(k).
\end{align}
This expression can be written in the simpler form
\[
\frac{\partial \mathcal{J}}{\partial W(k)_{ab}}
= -\sum_{i=1}^s r_{i,a}(k+1) z_{i,b}(k),
\]
where
\[
r_i(k) = \operatorname{diag}\left(\frac{d\sigma}{d v_{i,1}(k)},\dots,\frac{d\sigma}{d v_{i,n}(k)}\right)\lambda_i(k).
\]
From the derivative with respect to each entry of $W(k)$, the derivative with respect to the weight matrix $W(k)$ can be obtained
\begin{equation}\label{eq:dJdW}
\frac{\partial \mathcal{J}}{\partial W(k)}
= -\sum_{i=1}^s r_i(k+1)z_i(k)^{\top}=0,
\end{equation}
where $k \in \{2,\dots,\ell-1\}$. In principle, solving the stationarity conditions~\eqref{eq:dJdW} jointly with the state~\eqref{StateUpdate} and adjoint~\eqref{AdjointEq} equations would yield stationary points. However, because $z_i(k)$ and $\lambda_i(k)$, and hence $r_i(k)$, depend nonlinearly 
and recursively on $\{W(k')\}_{k'\le k}$  through the forward and backward passes, 
the resulting coupled nonlinear system is analytically intractable. Instead, it is customary to find the weights $W$ that enforce~\eqref{eq:dJdW} via gradient descent. This is done by viewing the loss functional as defining, for each fixed collection of weights
$W=(W(1),\dots,W(\ell-1))$, a scalar functional
\[
J(W) \stackrel{def}{=} \mathcal{J}(z(W),\lambda(W),W),
\]
where $z(W)$ and $\lambda(W)$ satisfy the forward~\eqref{StateUpdate} and backward~\eqref{AdjointEq} recursions. 
The gradient descent update for minimizing $J$ then takes the form
\begin{align}
W^{(q+1)}(k)
&=
W^{(q)}(k)
-\eta_q \frac{\partial \mathcal{J}}{\partial W(k)}\Big|_{W=W^{(q)}}\notag\\
\label{GradDesc}
&=
W^{(q)}(k)
+\eta_q \sum_{i=1}^s r_i^{(q)}(k+1)\,z_i^{(q)}(k)^T,
\end{align}
where $ k\in\{1,\dots,\ell-1\} $, the superscript $q$ 
denotes the $q$-th iterate of the gradient update,  $r_i^{(q)}(k)$ and $z_i^{(q)}(k)$ are obtained 
from the backward and forward passes using the current weights $W^{(q)}$, and $\eta_q$ is the step size also known as the learning rate.

Combining the necessary conditions for optimality with the gradient descent iterates results in the backpropagation algorithm that can be summarized as the iteration of the following steps:
\begin{enumerate}
\item \textbf{Forward pass:} propagate the input $z_i(1)$  to the output $z_i(\ell)$ using~\eqref{StateUpdate}.
\item \textbf{Backward pass:} initialize the adjoint $\lambda_i(\ell)$ at layer $\ell$ with~\eqref{FinalCondition} and backpropagate it to $\lambda_i(1)$ using~\eqref{AdjointEq}.
\item \textbf{Gradient descent:} update the weights $W$ using~\eqref{GradDesc}.
\end{enumerate}

\subsection{The modern connection}
\label{SSec:ModernConnection}

The connection between optimal control and backpropagation presented in~\emph{\nameref{SSec:ClassicalConnection}} consists in formulating the training of NNs as an optimal control problem in discrete-time, deriving  necessary conditions for optimality, and observing that those conditions essentially define backpropagation. As discussed in~\emph{\nameref{DiffEqs}}, for certain NN architectures, such as the ResNet, the state update equation can be regarded as the discretization of a differential equation. Similarly, the optimal control problem addressed in~\emph{\nameref{SSec:ClassicalConnection}} can be seen as the discretization of a continuous-time optimal control problem. This observation exposes an alternative path towards the training of NNs: formulate necessary conditions for optimality using an optimal control formulation in continuous-time and then discretize them to obtain a training algorithm. It is shown in this section that both approaches lead to the same solution, provided that suitable discretization techniques are employed. 

Redoing the analysis in~\emph{\nameref{SSec:ClassicalConnection}} for a ResNet requires changing the expanded loss functional $\mathcal{J}(z,W,\lambda)$ to
\begin{align*}
\sum_{i=1}^s L(y_i,\Phi(x_i))+\sum_{i=1}^s\sum_{k=1}^\ell& \lambda^T_i(k)\left[z_i(k)-z_i(k-1)\right.\notag\\
&\left.-w(k-1)\Sigma(W(k-1)z_i(k-1))\right]\notag
\end{align*}
in order to account for how the state is updated across layers in a ResNet, see~\eqref{ResNetStateUpdate}. The corresponding necessary conditions for optimality become

\begin{align}
z_i(k)&= z_i(k-1)+w(k-1)\Sigma(W(k-1)z_i(k-1))\notag\\
\label{DiscOpt2}
\lambda_i(k)&= \lambda_i(k+1)+w(k)\frac{\partial \Sigma}{\partial v}W(k)\lambda_i(k+1)\\
\label{DiscOpt3}
0&= -\sum_{i=1}^s r_i(k+1)z_i(k)^{\top}.
\end{align}

The very same conditions can be obtained starting from the continuous-time optimal control problem defined by the minimization of
\begin{equation}
\label{LossContOptimal}
\sum_{i=1}^s L(y_i,z_{c,i}(1)),
\end{equation}
where $z_{c,i}(1)$ denotes the solution of~\eqref{ODE} at time $1$ when starting from initial condition $x_i$. Since multiple solutions are used in the loss~\eqref{LossContOptimal} an ensemble control system consisting of $s$ identical copies of~\eqref{ODE} running in parallel and controlled by the same input $W_c$ is considered. The necessary conditions for optimality are given by Pontryagin's maximum principle~\cite{benning2019deep,li2018maximum}
\begin{align}
\label{ContOpt1}
\dot{z}_{c,i}(t)& =\frac{\partial H}{\partial \lambda_{c,i}}=\Sigma(W_c(t)z_{c,i}(t))\\
\label{ContOpt2}
\dot{\lambda}_{c,i}(t) & = -\frac{\partial H}{\partial z_{c,i}}=-\frac{\partial \Sigma}{\partial v}W_c(t)\lambda_{c,i}(t)\\
0 & =\frac{\partial H}{\partial W_c},\notag
\end{align}
where the Hamiltonian $H$ is
$$H(\lambda_{c,i}(t),z_{c,i}(t),W_c(t)) = \sum_{i=1}^s \lambda_{c,i}^T(t)\Sigma(W_c(t)z_{c,i}(t)).$$
A naive discretization of~\eqref{ContOpt2}, using the forward Euler method, \emph{does not} lead to~\eqref{DiscOpt2}. In order to recover~\eqref{DiscOpt2}, the symplectic version of the forward Euler method, also known as the semi-implicit Euler method~\cite[Eq. (1.9)]{HaLuWa2006}, must be employed. It consists of the forward Euler method applied to~\eqref{ContOpt1} but requires discretizing~\eqref{ContOpt2} using the implicit rule
$$\lambda_{i}(k+1)=\lambda_{i}(k)-w(k)\frac{\partial \Sigma}{\partial v}W(k)\lambda_i(k+1),$$
which is simply~\eqref{DiscOpt2}.

Although the interplay between backpropagation and optimal control in discrete-time already allows for the migration of ideas from control to deep learning, the linkage with continuous-time optimal control opens new perspectives. Continuous-time control theory is substantially more developed than its discrete-time counterpart. Moreover, continuous-time results are often more intuitive thus making it easier to adapt/generalize to the setting of deep learning. Some examples of this migration of ideas are discussed in the next section.

\subsection{New training techniques}
\label{SSec:NewTraining}

The previous sections shed light into the interplay between optimal control and the training of NNs. However, the use of gradient descent is but one possible path to find the weights enforcing equality~\eqref{DiscOpt3}. When the weights are real valued,~\cite{li2018maximum} discusses why gradient descent may be preferable to directly solving~\eqref{DiscOpt3}. The work in~\cite{pmlr-v80-li18b} uses the extra generality afforded by the maximum principle to derive training algorithms for NNs whose weights belong to finite sets, in which case gradient descent is not an option.  Extending the optimal control perspective from one to two players brings us to the setting of games which proved useful to develop more efficient adversarial training algorithms in~\cite{NEURIPS2019_812b4ba2}, see also~\cite{pmlr-v120-seidman20a} for a rigorous convergence analysis. The modern continuous-time perspective was exploited in~\cite{benning2019deep} to obtain training algorithms where the discretization time step is regarded as one additional control input used for more efficient training.


Rather than linkages to optimal control, other advances in training techniques leveraged the view of NNs as discretizations of differential equations.  The important property of reversibility, addressed in~\cite{ChMeHaRuBeHo2018}, is intimately tied to training, since backpropagation requires both forward and backward passes. In particular, it shows how reversibility can be exploited for more efficient training and proposes new reversible architectures. Another important property is stability~\cite{HaRu2017,ChMeHaRuBeHo2018}, understood as Lipschitz continuity of the map sending the state of the first layer to the state of the last layer. Stability guarantees robustness of the network against noisy inputs or perturbations of the data. As discussed in~\cite{CeEhEtMcOwScSh2021}, the Lipschitz constant depends on the activation function $\sigma$ and the weights for the continuous-time model~\eqref{ODE}. For the discrete-time model \eqref{ResNetStateUpdate}, it also depends on the number of layers. The work in~\cite{CeEhEtMcOwScSh2021} discusses several architectures inspired by ideas from dynamical systems, such as Hamiltonian dynamics, that ensure stability by construction rather than relying on training.


\color{black}

\section{Universal approximation}
\label{Sec:UniversalApproximation}

One of the reasons underlying the wide applicability of NNs is their perceived ability to approximate arbitrary functions. This is known as the power of universal approximation. To make this notion precise, recall that a NN is a family of functions $\Phi_W:\R^{n_1}\to \R^{o}$ parameterized by the weights $W$, see~``\nameref{WhatAreNNs1}'' sidebar. It can then be asked if, given a function $\psi:K\to \R^{o}$ defined on a domain of interest $ K\subseteq \R^{n_1}$, there is a choice of weights $W$ so that $\Phi_W$ approximates $\psi$ arbitrarily well with respect to a chosen topology (uniform, $L^p$, Sobolev, etc.).

\subsection{A concise historical account of universal approximation in neural networks}

Classical universal approximation results established in late '80s and early '90s focused on proving that feedforward NNs with small depth, that is, a small number of layers, but unrestricted width, that is, an arbitrary number of neurons per layer, can approximate continuous functions on compact sets, under mild  assumptions on the activation functions. In particular, universality of sigmoidal activation functions was proved in~\cite{GC:89,UApprox89}. In a series of follow-up work, including~\cite{KH:91,AP:99}, it was shown that for broad classes of NNs, \emph{non-polynomial} activations are essentially necessary and sufficient for approximating continuous functions on compact sets.  The mathematical machinery behind these results is functional analysis' density arguments. A different argument was employed in~\cite{256500} to establish that the approximation error decays at the rate $1/n$, where $n$ is the number of sigmoids in the hidden layer, independently of the dimension of the approximated function's domain.

The modern deep learning era revived these questions due to the evolution in NN architectures. Deep NNs have large depth, that is, a large number of layers, but bounded width, that is, a bounded number of neurons per layer. This mismatch between architectures prevents a direct application of the classical universal approximation results. Establishing universal approximation for deep neural networks relies on understanding composition of maps, which, due to nonlinearities of the activation functions, is challenging. 

The literature on this subject is quite recent. One of the early results,~\cite{OneNeuron}, focused on the class of ReLU activation functions to establish that one neuron per layer suffices to approximate an arbitrary Lebesgue integrable function from $\R$ to $\R$ with respect to the $L^1$ norm. The results that followed addressed more general functions from $\real^{n_1}$ to $\real^{o}$. For instance, when using the ReLU activation function and $o=1$, it was shown in~\cite{OneNeuron} that universal approximation with respect to the $ L^1$ topology can be achieved with a minimum width between $n_1+1$ and $n_1+4$. The theory of universal approximation for deep NNs substantially advanced in early 2020, with notable discoveries in~\cite{park2021minimum}, and particularly in~\cite{PK-TL:20}, where universal approximation was established with respect to the \emph{infinity norm} for non-residual deep NNs, allowing for  general classes of activation functions. More precisely, it is shown in~\cite{PK-TL:20} that under very mild conditions on the activation functions, any continuous function $ \psi: K \rightarrow \real^{o} $, where $ K \subset \real^{n_1} $ is compact, can be approximated in the infinity norm using a deep NN of width $ n_1+o+2 $. 

\subsection*{The control theory vantage point} 
\label{Ssec:ControlTheoryVantage}
Inline with the objectives of this article, this section provides the control-theoretic viewpoint on universal approximation developed in~\cite{PT-BG:21,PT:BG:22-tac,PT-BG-24-c}, and also in~\cite{agrachev2022control}. The starting point is the continuous limit of a ResNet, as explained in~\emph{\nameref{DiffEqs}}, and written as
\begin{equation}
\label{ControlSystem}
\dot{z}_c(t)=w_c(t)\Sigma(W_c(t)z_c(t)).
\end{equation}
Note that~\eqref{ControlSystem} differs from~\eqref{ODE} by the presence of $w_c(t)$. This is achieved by writing the weight $w(k)$ in~\eqref{ResNetStateUpdate} as the product $w(k)=\tau\, w_c(k)$ and letting $\tau\to 0$ as $\ell\to \infty$ so that $w_c(t)$ is the continuous limit of $w_c(k)$. For notational convenience, denote by $\gamma:\R^n\to\R^n$ the \emph{flow} of~\eqref{ControlSystem}, that is, the map sending an initial condition $z_{c}(0)$ to the point reached by the solution of~\eqref{ControlSystem} at time $t=1$ under the input $(w_c,W_c)$.  

The objective of approximating $\psi:K\to \R^{o}$ with $\gamma$ requires the existence of control inputs $(w_c,W_c)$ for which the corresponding flow $\gamma$ satisfies $\gamma(x_i)=\psi(x_i)$ for every sample point $x_i\in K\subseteq \R^{n_1}$ or, at least, that $\gamma(x_i)$ is close to $\psi(x_i)$. This can be understood as a controllability problem for which the initial conditions $x_1,\hdots,x_s$ have to be steered to $y_1=\psi(x_1),\hdots,y_s=\psi(x_s)$ at time $t=1$ by the same open-loop control input $(w_c,W_c)$.

At this point, it is worth pointing out the sharp difference between this setting and earlier work on controllability of differential equation models for neural networks, for example,~\cite{SONTAG1999121}, where the weights were assumed to be constant and an exogenous control signal was fed into the neurons.  

The simultaneous control of multiple initial conditions is naturally related to the classical settings of ensemble control theory, for example,~\cite{LJS-KN:06,brockett2007optimal}. However, in ensemble control, the motion of each member of the ensemble is governed by a different, even if very similar, control system. In the setting of universal approximation, the motion from each initial condition is governed by the same control system under the same input. This setting  is most closely related to~\cite{AA-AS:20}, where controllability results for ensembles of infinitely many copies of the same control system are considered. With this perspective in mind, there are two key problems that can be considered. 

\emph{{\bf Memorization Problem:} Let $\psi:\R^{n_1}\to \R^{o}$ be a continuous function, $K_{\text{samples}}\subset \R^{n_1}$ be a finite set, and $\varepsilon\in \R^+$ be the approximation accuracy. Given any desired $\varepsilon$ and $p\in [1,+\infty]$, does there exist an initialization map $f:\R^{n_1}\to \R^n$, a finalization map $h:\R^n\to \R^{o}$,
and an input $(w_c,W_c):[0,1]\to \R\times\R^{n\times n}$
so that the flow $\gamma:\R^n\to \R^n$ defined by the solution of~\eqref{ResNetStateUpdate}
under the said input satisfies
\begin{equation*}
\bigl\| \psi-h\circ \gamma\circ f \bigr\|_{L^p(K_{\text{samples}})}\le \varepsilon\, ?   
\end{equation*}}

The notation $\Vert\cdot\Vert_{L^p(K_{\text{samples}})}$ denotes the $L^p$ norm on the set $K_{\text{samples}}$, in other words, the approximation guarantee only needs to hold for elements in the finite set $K_{\text{samples}}$. The purpose of the initialization function $f$ is to map $\R^{n_1}$ to $\R^n$ since the continuous model~\eqref{ControlSystem} was developed assuming all the layers have the same number $n$ of neurons. Similarly, the finalization map converts $\R^n$ to $\R^{o}$.


We note that the memorization problem above is a simultaneous controllability problem which is difficult to solve since the \emph{same} differential equation~\eqref{ControlSystem} governs the evolution of the state from each initial condition $z_{c}(0)$ and under the \emph{same} open-loop control input.

The next problem considers the more challenging case of universal approximation on compact sets. 

\emph{{\bf Universal Approximation Problem:} Let $\psi:\R^{n_1}\to \R^{o}$ be a continuous function, $K\subset \R^{n_1}$ be a compact set,  and $\varepsilon\in \R^+$ be the approximation accuracy. Given any desired $\varepsilon$ and $p\in [1,+\infty]$, does there exist a continuous initialization map $f:\R^{n_1}\to \R^{n}$,
a continuous finalization map $g:\R^{n}\to \R^{o}$, and an input
$(w_c,W_c):[0,1]\to \R\times\R^{n\times n}$
such that the flow $\gamma:\R^n\to \R^n$ defined by the solution of~\eqref{ControlSystem}
under the said input satisfies
\begin{equation*}
\bigl\| \psi- h\circ \gamma\circ f \bigr\|_{L^p(K)}\le \varepsilon \, ?     
\end{equation*}}

 \begin{figure}[htb!]
   \centering 
\includegraphics{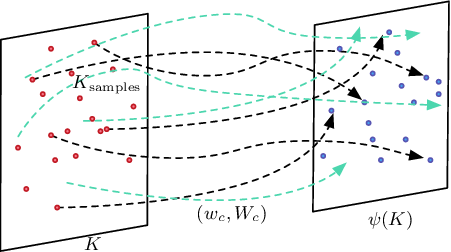}
     \caption{A schematic picture of the two problems under study. The objective in the memorization problem is to find a control input 
     $ (w_c,W_c) $ that takes all the sample points in $ K_{\text{samples}} $ (represented by red dots) to their corresponding points in $ \psi(K_\textrm{samples}) $  (represented by blue dots); some of these assignments are shown by black dashed arrows. In the universal approximation problem, the points outside $ K_{\text{samples}} $ also need to be carefully mapped to guarantee approximation in the uniform sense; some of these assignments are represented by dashed green arrows.  
     }\label{fig_ua}
\end{figure}

Contrary to the memorization problem, the universal approximation problem requires the approximation guarantees to hold for every element in $K$, which is no longer a finite set. Figure~\ref{fig_ua} provides a schematic picture of memorization and universal approximation. 
Naturally, geometric control theory plays a key role in the study of the  memorization problem when framed as a controllability problem. In particular, a keen reader would realize that the classical Lie algebraic conditions for accessibility require the derivatives of the activation function to appear in the analysis of this problem. As shown in~\cite{PT:BG:22-tac}, a sufficient condition for controllability is provided by asking the activation function $\sigma$, or its derivatives, to satisfy an ordinary differential equation with \emph{quadratic righthand side}; interestingly, a large class of widely used activations functions satisfy this condition.




The key result in~\cite{PT:BG:22-tac} provides answers to both memorization and universal approximation problems. In particular, given a continuous function $\psi: \R^{n_1}\to \R^{o}$,  
    and assuming that the activation function satisfies a differential equation with quadratic right-hand side, the following was established: 
    \begin{itemize}
        \item {\bf Memorization:} The memorization problem has a solution for any $\varepsilon\in \R^+$. 
        
        \item {\bf Universal Approximation:} The problem has a solution for any $\varepsilon\in\R^+$ when $n$ is given by $2\max\{n_1,o\}~+~1 $.         
    \end{itemize}  
The results in~\cite{PT:BG:22-tac} are, in fact, sharper in the sense that, under suitable technical assumptions, the memorization problem can be solved with $\varepsilon=0$ for almost all data sets and universal approximation can be achieved using a smaller number of neurons for a more restricted class of functions.  Notably, in the case where $ n_1=o$, the universal approximation statement above improves upon the one in~\cite{PK-TL:20}, which requires $n_1+o+2 $ neurons. 


The problem of universal approximation led to a fruitful bridge between control theory and deep learning, see for instance~\cite{kvalheim2025autoencodingdynamicstopologicallimitations,bayram2025geometric,veeravalli2023nonlinear,hanson2025fitting,hanson2025expressivity}. In particular, the composition of flows, a concept with deep roots in control theory, has become instrumental for the study of Neural ODEs. This perspective has spurred substantial activity in the past few years at this intersection, including~\cite{agrachev2024generic,scagliotti2021deep,scagliotti2025normalizing,cheng2025interpolation,10999048,alvarez2024interplay,prieur2025unified}.

We close this section by discussing a few intriguing directions related to the approximation power of Neural ODEs for density estimation by analyzing controllability properties of the corresponding continuity equation, where the driving vector field is parameterized by a Neural ODE. For instance, results in~\cite{ruiz2023neural,KE-BG-AB-SO:22} establish approximate controllability of the controlled continuity equation on the space of probability measures, with connections to optimal transportation theory: given initial and target distributions, it is shown that the Neural ODE's weights can be chosen so that the resulting terminal distribution is arbitrarily close to the target in an appropriate measure.  The literature on this subject is fast growing~\cite{elamvazhuthi2023learning,ruiz2024control,geshkovski2022turnpike}, and the techniques have also recently been  employed for continuity equations corresponding to the self-attention dynamics in transformers~\cite{geshkovski2024measure,DA-BG:24}.




\color{black}
\section{Large language models, transformers, and consensus}
\label{Sec:LLMsTransformesConsensus}

Large language models (LLMs) is now used as a generic term for NN architectures that generate textual responses to user's queries in natural language. It was formed by pre-pending ``large'', referring to model and training data size, to ``language models'', referring to the models used in natural language processing. Transformer-based LLMs  are a good example of how ideas from control can be used to better understand deep learning models. First, by passing to the continuous limit, as explained in~\emph{\nameref{DiffEqs}}, the transformer architecture can be described by a differential equation. Second, the differential equation model can be interpreted as a generalization of the consensus models studied in control theory. This section explains these connections and how ideas from consensus on manifolds reveal the role of the attention mechanism in transformer-based LLMs. 

An LLM first converts the query or prompt into a sequence of tokens $z_1(1),\hdots,z_m(1)$ using the map $f$ appearing in~\eqref{TransformerEq}. Then, a sequence-to-sequence NN is employed to process the sequence of tokens and produce the output sequence $z_1(\ell),\hdots, z_m(\ell)$. Although there are several choices for this NN, the most widely used is the transformer described in the sidebar ``\nameref{WhatAreNNs2}''. Other possibilities are the RWKV model~\cite{LiXiChWu2025}, which is based on a recurrent neural network (RNN), and the Mamba model, which is based on a selective state-space-model (SSM). See sidebar ``\nameref{WhatAreNNs3}'' for an introduction to RNNs and SSMs. The output of the sequence-to-sequence model, $z_1(\ell),\hdots,z_m(\ell)$, is processed by the output map $h$, see~\eqref{TransformerEq}, to predict the next token $z_{m+1}(1)$ which is appended to the end of the input token sequence to generate the longer sequence $z_1(1),\hdots, z_m(1),z_{m+1}(1)$. This longer sequence is then processed by the sequence-to-sequence NN to predict the next token, analogously to a feedback loop. The process is iterated until the reply sentence or sentences are completed.

\subsection{Self-attention dynamics}

Denote by
\begin{equation*}
Z(k)=\begin{bmatrix}z_1(k)\vert z_2(k)\vert\hdots\vert z_m(k)\end{bmatrix}\in\R^{d\times m}
\end{equation*}
the matrix representation of the tokens at layer $k$ of a transformer NN. The transformer block updates each token according to the recurrence formula
\begin{align}\label{TransformerBlock1}
\tilde z_i(k) & =\mathbf{n}(z_i(k)+A_i(Z(k))),\\\label{TransformerBlock2}
z_i(k+1) & =\mathbf{n}(\tilde z_i(k)+r_k(\tilde z_i(k))),
\end{align}
where $\mathbf{n}:\mathbb R^d\to\mathbb R^d$ is the normalization map, $r_k(\tilde z_i(k))=W_2(k)\Sigma(W_1(k)\tilde z_i(k))$, and
\begin{equation*}
A=\begin{bmatrix}A_1\vert A_2\vert\hdots\vert A_m\end{bmatrix}:\R^{d\times m}\to\R^{d\times m}
\end{equation*}
is the self-attention mechanism (see sidebar ``\nameref{WhatAreNNs2}'' for a detailed description of these objects).

Focusing on the self-attention mechanism by removing the skip-connection~\eqref{TransformerBlock2}, the transformer block \eqref{TransformerBlock1}, \eqref{TransformerBlock2} simplifies to
\begin{align}\label{AttentionBlock}
z_i(k+1)=\mathbf{n}(z_i(k)+A_i(Z(k))).
\end{align}

It was first noted in \cite{LuLiHeSuDoQiWaLi2019} and further developed in \cite{DuGaChCh2021} that the attention block \eqref{AttentionBlock} resembles the dynamics of a multi-particle system and, following the ideas presented in~\emph{\nameref{DiffEqs}}, it can be modeled by a system of differential equations. 

In the differential equation model, the matrix representation of tokens is parameterized by $t\in\R_0^+$:
\begin{equation*}
Z_c(t)=\begin{bmatrix}z_{c,1}(t)\vert z_{c,2}(t)\vert\hdots\vert z_{c,m}(t)\end{bmatrix}\in\mathbb R^{d\times m}.    
\end{equation*}
Note the deviation from the description in~\emph{\nameref{DiffEqs}}, where the time domain was normalized to $[0,1]$. This section discusses  the asymptotic behavior of transformers thus suggesting $\R_0^+$ as the natural time domain. As it will be seen, existence of solutions is not a concern since the state space will be a product of spheres, hence compact. Similarly to the token matrix, the weight matrices are functions of time $W_{Q_j}(t),W_{K_j}(t)\in\mathbb R^{e\times d}$ and $W_{V_j}(t)\in\mathbb R^{e'\times d}$. Hence, the attention mechanism can be written as
\begin{align}\nonumber
A_c(t,Z_c) & =\begin{bmatrix}A_{c,1}(t,Z_c)\vert A_{c,2}(t,Z_c)\vert\hdots\vert A_{c,m}(t,Z_c)\end{bmatrix}\\\label{Attention_Continuous}
& =\sum_{j=1}^a W_{U_j}(t)\,Z_c\,\mathrm{softmax}\!\left(\frac{Z_c^\top W_{P_j}(t)Z_c}{\sqrt{e}}\right),
\end{align}
where $a$ is the number of heads, the softmax is applied column-wise, and the matrices
\begin{align*}
W_{U_j}(t) & =W_{o_j}W_{V_j}(t)\in\mathbb R^{d\times d},\\
W_{P_j}(t) & =W_{K_j}^\top(t)W_{Q_j}(t)\in\mathbb R^{m\times m}
\end{align*}
were used for brevity. When there is only one head ($a=1$), the subscript $j=1$ is omitted for simplicity, that is, $W_{U_1}(t)$ simplifies to $W_U(t)$ and $W_{P_1}(t)$ simplifies to $W_P(t)$.

The self-attention mechanism~\eqref{AttentionBlock} can now be written
\begin{align}\label{AttentionODE}
\dot z_{c,i}=T_{z_{c,i}}\mathbf{n}\cdot A_{c,i}(t,Z_c(t)),
\end{align}
where $T_z \mathbf{n}:\mathbb R^d\to\mathbb R^d$ denotes the tangent map of $\mathbf{n}$ at $z\in\mathbb R^d$ using the identification $T_z\mathbb R^d\simeq\mathbb R^d$.

For the purpose of  analyzing the asymptotic behavior of $z_{c,i}$, the normalization map is typically chosen to be the projection onto the $(d-1)$-dimensional sphere $\mathbb S^{d-1}=\{z\in\mathbb R^d|z^\top z=1\}$. In other words,
\begin{equation}\label{normalization}
\mathbf{n}(z)=\frac{z}{\Vert z\Vert},    
\end{equation}
where $\Vert z\Vert$ denotes the Euclidean norm of $z\in \R^d\backslash\{0\}$ thus forcing tokens to have unit norm. Therefore, the configuration space of both the discrete \eqref{AttentionBlock} and the continuous \eqref{AttentionODE} dynamical systems governing the evolution of  attention is the product of $m$ copies of the $(d-1)$-sphere
\begin{equation*}
(\mathbb S^{d-1})^m=\underbrace{\mathbb S^{d-1} \times{\hdots}\times\mathbb S^{d-1}}_{m\text{-times}}.
\end{equation*}

Recall that the tangent space of the sphere at each point $z\in\mathbb S^{d-1}$ consists of all vectors that are orthogonal to $z$, that is
\begin{equation*}
T_z\mathbb S^{d-1}=\{X\in\mathbb R^d|z^\top X=0\}.
\end{equation*}
The tangent map of the projection thus reads
\begin{equation*}
T_z \mathbf{n}\cdot X=(\mathbb I_d-z z^\top)\cdot X
\end{equation*}
for each $X\in\mathbb R^d$.

In real transformers, however, the standard layer normalization,
\begin{equation*}
\mathbf{n}(z)=\frac{1}{\sigma(z)}(z-\mu(z)\mathbf 1)\star\gamma+\beta,
\end{equation*}
is the most common choice. In the previous expression, $\mathbf 1$ denotes the vector $(1,\hdots,1)\in\mathbb R^d$, $\star$ denotes the element-wise product of vectors, and $\gamma,\beta\in\mathbb R^d$ are the \emph{learned scale} and \emph{shift}, respectively. Similarly, $\mu(z)$ and $\sigma(z)$ denote the mean and standard deviation of $z$, respectively.
An easy check shows that, if $\gamma=\gamma_0\mathbf 1$ for some fixed $\gamma_0\neq 0$, then the tokens lie on the $(d-1)$-sphere of center $\beta$ and radius $d\gamma_0^2$. Therefore, the theoretical results with the normalization to the sphere \eqref{normalization} also apply to the standard layer normalization. For arbitrary scale parameters, it is conjectured that the tokens also lie on a certain hypersurface of $\mathbb R^d$ diffeomorphic to the $(d-1)$-sphere, making the theoretical results still valid.

\subsection{Consensus dynamics}
\label{Ssec:Consensus}

Many empirical works starting with \cite{DoCoLo2021} observed that tokens tend to converge to consensus as the depth of the transformer increases, that is, the model collapses as all tokens converge to a single point. The first theoretical explanation for this observed behavior was provided in~\cite{BG-CL-YP-PR:23,BG-CL-YP-PR:24}. The  continuous attention dynamics \eqref{AttentionODE} was regarded as a mean-field interacting particle system and it was shown that, for almost every initial condition, all tokens converge to a fixed position as time goes to infinity. The analysis was restricted to a simplified model with only one head, $a=1$, and weight matrices assumed to be constant and proportional to the identity, $W_{U}(t)=\mathbb I_d$ and $W_{P}(t)=\beta\,\mathbb I_m$, where $\beta>0$ is known as the \emph{coldness} or \emph{inverse temperature}.

Under different assumptions on $d$, $m$ and $\beta$, asymptotic convergence to consensus was proven using techniques from partial differential equations and ergodic theory. In subsequent papers \cite{GeKoPoRi2024,BrPaAg2025,Ri2026}, it was also shown that metastability holds under certain conditions on the parameters and the initial conditions, that is, tokens first form several clusters and remain there for some time before eventually collapsing to a single one. The aforementioned papers address the full self-attention mechanism, but analogous results have been proven for the \emph{causal} (auto-regressive) attention \cite{Ri2026}. The causal attention mechanism is a modification of \eqref{Attention_Continuous} where the attention for the $i$-th token only depends on the previous ones.

Ideas from consensus dynamics and control theory have also demonstrated usefulness in the asymptotic analysis of the transformer dynamics \cite{RoSiTa2024,RoSiTa2025,Al2025}, enabling the analysis of \eqref{AttentionODE} for more general weight matrices.  To better understand the relation between~\eqref{AttentionODE} and the consensus models used in control, note that~\eqref{AttentionODE} can also be written in the following form when there is a single head ($a=1$):
$$\dot{z}_{c,i}=T_{z_{c,i}}n\cdot W_U(t)\sum_{j=1}^m a_{ij}(t,Z_c)z_{c,j},$$
where the coefficients $a_{ij}$ appearing in the above linear combination are strictly positive and sum to $1$, that is, $\sum_{j=1}^m a_{ij}=1$. Differently from most  consensus models investigated in control, the coefficients $a_{ij}$ are nonlinear functions of the state $Z_c$. The presence of the matrix  $W_U(t)$ is another difference with respect to the consensus models in the control literature.

For causal self-attention, an input-to-state stability argument shows that the behavior of the system is determined by the evolution of the first token. Causality results in a triangular structure on the dynamics suggestive of control systems in strict feedback form~\cite{RNCD}. Moreover, the dynamics of token $i$ is input-to-state stable with respect to distance between tokens $1,\hdots,i-1$ and the consensus set, that is, the set defined by $z_{c,1}=z_{c,2}=\hdots=z_{c,i-1}$. Hence, asymptotic stability towards the consensus set is established using input-to-state stability of cascade interconnections.

The results proven using these techniques point in the same direction as previous contributions to this problem: the model collapses as all tokens converge to consensus as time increases. The main advantage of using control theoretic techniques compared to the mean-field approach is that they allow for addressing non-identity  and time-varying weight matrices. This situation is significantly more realistic as weights depend on the layer in real transformers. 
Namely, the results in~\cite{RoSiTa2024} for the causal self-attention case may be summarized in three cases:
\begin{enumerate}
    \item $a\geq 1$ and $W_{U_j}(t)=\mathbb I_d$ for all heads. Then the first token is fixed and, thus, all tokens converge to the initial position of the first token. The domain of attraction of this consensus is co-null, that is, its complement has zero measure.
    \item $a=1$ and $W_{U}(t)=W_U$ constant and symmetric. Then the first token converges to the principal eigenvector of $W_U$, that is, the eigenvector (up to sign) corresponding to maximum eigenvalue. The sign is determined by the initial position of the first token. More recently,~\cite{Al2025} extended this result to the full self-attention. They also identified new types of equilibria showing that all of them are unstable.
    \item $a=1$ and $W_{U}(t)$ bounded and symmetric for all times. Then the principal eigenvalue of $W_{U}(t)$ also depends on time. As a result, tokens do not converge to a fixed position, but they keep following the principal eigenvector instead. Specifically, they converge to a ball around the principal eigenvector forming a moving cluster. The size of the ball depends on how fast the eigenvector moves.
\end{enumerate}

Even though a fully general asymptotic analysis of the model \eqref{AttentionODE} is still missing, all the results to date suggest model collapse. 

\subsection{Empirical validation of theoretical analysis}
The collapse of the attention mechanism, explained by the theoretical analysis in~\cite{RoSiTa2024}, is  supported by empirical results on LLMs that include the skip-connection~\eqref{TransformerBlock2} and thus go beyond the simplifying assumption made when deriving the theoretical differential equation model.

Several tests on the GPT-2 XL and the GPT-Neo 2.7B models are reported in~\cite{RoSiTa2025}. Since the theoretical results are asymptotic, the depth was increased by looping the transformers, that is, by feeding the tokens at the final layer as inputs to the first layer. 

The error function
\begin{equation}
\label{E}
    E(k)=  1-\frac{1}{m}\sum_{i=1}^m \frac{z_1^\top(k)\,z_i(k)}{|z_1(k)|\,|z_i(k)|}
\end{equation}
was used to measure how far the tokens are from the first one. Note that $E(k)\geq 0$ and it vanishes only when all tokens are aligned, $z_1(k)=z_2(k)\hdots=z_m(k)$, that is, when consensus is reached. The average of \eqref{E} was computed across 100 random prompts generated by uniformly sampling 200 tokens from the GPT-2 tokenizer's vocabulary.

\begin{figure}[!b]
    \centering{
            \includegraphics[width=0.45\textwidth, keepaspectratio]{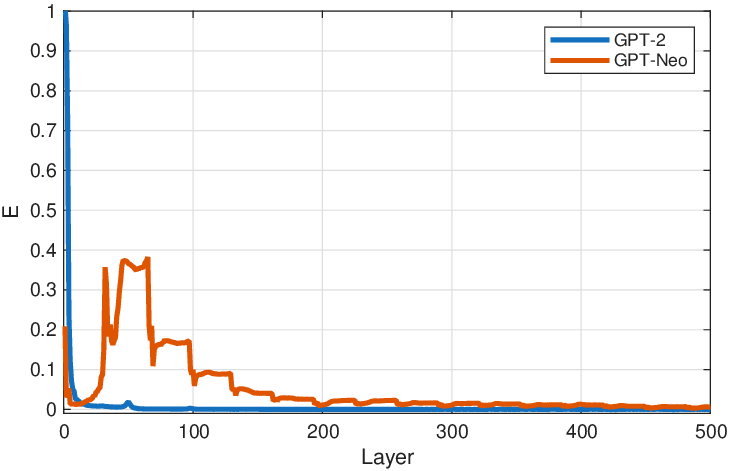}}
    \caption{Comparison between the GPT-2 XL and GPT-Neo 2.7B architectures with feedforward layers removed; evaluation of the average of~\eqref{E} across all the random prompts.}
    \label{pretrained-attention}
\end{figure}

\begin{figure}[!b]
    \centering{
        \includegraphics[width=0.45\textwidth, keepaspectratio]{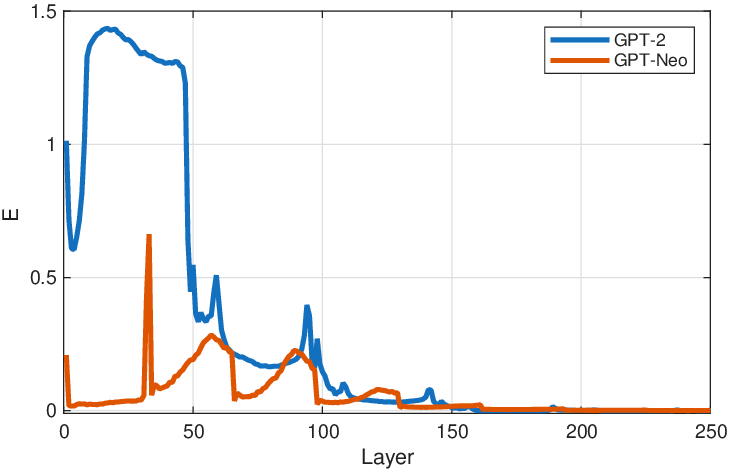}}
    \caption{Comparison between the GPT-2 XL and GPT-Neo 2.7B architectures with the full model; evaluation of the average of~\eqref{E} across all the random prompts.}
    \label{pretrained-full}
\end{figure}

\begin{figure}[!b]
    \centering{
        \includegraphics[width=0.45\textwidth, keepaspectratio]{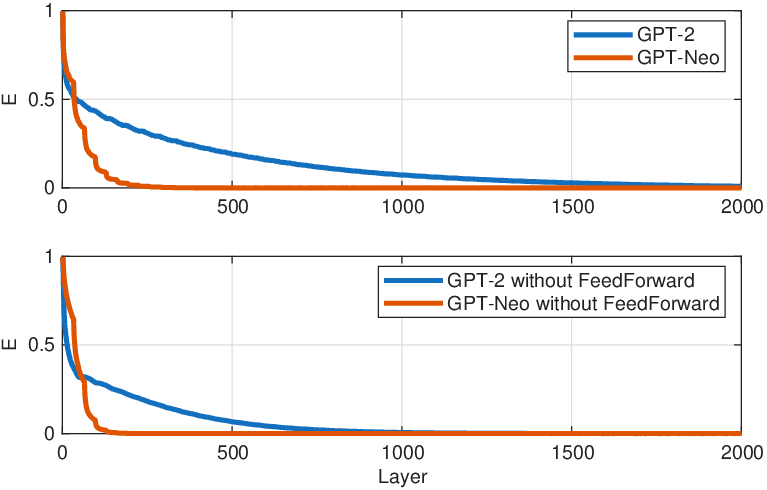}}
    \caption{Comparison between the GPT-2 XL and GPT-Neo 2.7B architectures with fixed and randomly chosen weight matrices. Each model was evaluated with and without feedfoward layers using the average of~\eqref{E} across all the random prompts.}
    \label{random-fixed}
\end{figure}

\begin{figure}[!b]
    \centering{
        \includegraphics[width=0.45\textwidth, keepaspectratio]{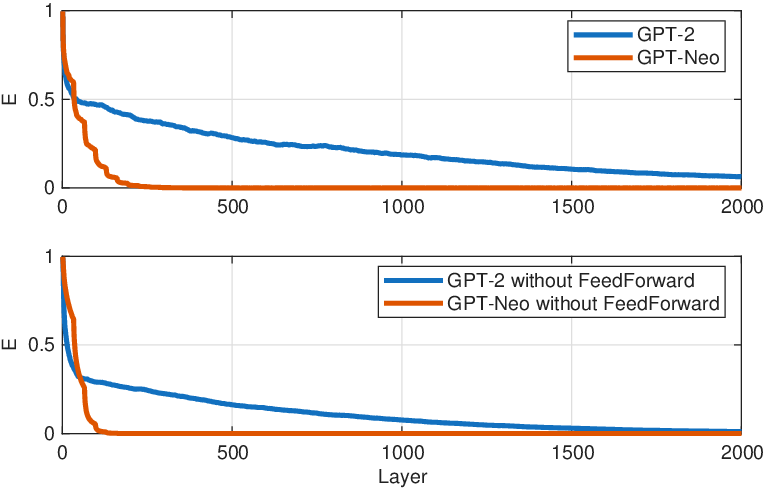}}
    \caption{Comparison between the GPT-2 XL and GPT-Neo 2.7B architectures with random weight matrices chosen before each model pass. Each model was evaluated with and without feedfoward layers using the average of~\eqref{E} across all the random prompts.}
    \label{random-new}
\end{figure}

Two types of experiments comparing the GPT-2 XL and the GPT-Neo 2.7B models were performed (all figures are extracted from \cite{RoSiTa2025}):
\begin{enumerate}
    \item \emph{Evolution of tokens using pre-trained LLMs.} Using the  pre-trained weights provided by the Hugging Face
library \cite{wolf2020transformers}.  Looping the transformers yields weight matrices that are time-varying but periodic. Figure \ref{pretrained-attention} shows the evolution of $E$ when the skip-connection~\eqref{TransformerBlock2} is absent and thus there is agreement between the theoretical model and the LLMs. Figure \ref{pretrained-full} shows the same experiment with the full model, that is,~\eqref{TransformerBlock1} and~\eqref{TransformerBlock2}.
    \item \emph{Evolution of tokens using random weight matrices.} As for the previous case, the experiments were carried out for the self-attention only, as well as for the full model. In Figure \ref{random-fixed}, the random weights were fixed, that is, they were the same for each pass of the model, providing time-varying, periodic weights. In Figure \ref{random-new}, new random weights were generated before each pass of the model, providing time-varying, non periodic weight matrices.
\end{enumerate}

We can observe that in all the above described cases the function $E$ converges to zero thus showing that all the tokens converge to the same value. The experiments with randomly generated matrices suggest that consensus is a structural property of the architecture and does not depend on the choice of weights or prompts. It can be observed that the rate of convergence depends on the skip-connections: for the GPT-2 model it increases, whereas for the GPT-Neo model it decreases. Nevertheless, consensus still takes place when the skip-connection layers are considered.

As an illustration of consensus, we took the first sentence of this article, \texttt{Deep neural networks, in their different shapes and flavors, are having a remarkable impact in almost every area of science and engineering.}, and passed it through GPT-2-XL. The decoded output is presented in Table~\ref{Tab:Sentences} along with the corresponding number of passes. We can appreciate that after 4 passes consensus has been reached since all the tokens are close enough so as to decode to the same word: \texttt{The}.

\begin{table}
\caption{Illustration of consensus by decoding the output produced by passing the same sentence through GPT2-XL an increasing number of times}
\begin{tabular}{c|l}\label{Tab:Sentences}
\# passes & Output\\
\hline    
1 & \texttt{ly networks are which particular}\\
& \texttt{simplest forms and sizes, are the a }\\
&\texttt{moment impact on many every field }\\
&\texttt{of science and technology. that }\\
&\texttt{automated 
them is used.' produce They}\\
\hline 
2 & \texttt{, neural for network network  
}\\
&\texttt{the the andy the the same}\\
&\texttt{ good good, The a the,,,,, a, }\\
\hline
4 & \texttt{TheTheTheTheTheTheTheTheThe}\\
&\texttt{TheTheTheThe"TheTheTheTheThe}\\
&\texttt{TheTheTheTheTheTheThe"}\\
\hline
10&\texttt{TheTheTheTheTheTheTheTheThe}\\
&\texttt{TheTheTheTheTheTheTheTheThe}\\
&\texttt{TheTheTheTheTheTheTheThe}\\
\end{tabular}
\end{table}

\section{Other research directions}

This section focuses on more recent and innovative work that opens exciting new directions at the interplay between control theory and deep learning. Given the recent and exploratory nature of these developments, the discussion is less detailed and emphasizes the key ideas behind each approach, their potential advantages, and how they may contribute to the broader landscape of deep learning.

\subsection{Selective State Space Models}
RNNs and SSMs are the architectures that most closely resemble the models studied in control theory as they are directly given as discrete-time nonlinear control systems, see~``\nameref{WhatAreNNs3}'' sidebar. RNNs are statefull models and thus can produce outputs that depend on past inputs since input information is stored in the state. In principle, this makes these models especially useful when there are long range dependencies between inputs and outputs. In practice, however, RNNs are difficult to train due to the vanishing (or exploding) gradient problem. In contrast, transformers can only reason about dependencies that occur in a window whose length corresponds to the number of processed tokens. Yet, this restriction makes them more reliable, parallelizable, and scalable.

Much excitement exists around SSMs; they are also statefull models yet overcome the complexity limitations brought by the difficulty of parallelizing the training of RNNs. The training and inferencing of SSMs is parallelizable, since the matrix  $A(k,z(k,i))$ is restricted to be diagonal, but this restriction also raises the question of understanding if it compromises the ability of SSMs to model long range dependencies between inputs and outputs. Answering this question is still an active area of research. It is established in~\cite{nishikawa2024state} that SSMs are comparable to transformers in their ability to identify the most relevant information within a text. It is also known, see~\cite{merrill2024illusion}, that there are simple state tracking tasks that cannot be accomplished by SSMs, neither by transformers, but can be performed by RNNs. 

Another question that remains open is determining which tasks are better suited for transformers or for SSMs. A recent study~\cite{waleffe2024empirical} conducted a large-scale empirical comparison between Mamba and Mamba-2 models, and transformers. The results show that while Mamba-2 can outperform transformers of comparable size, this advantage is not transversal across all tasks. Interestingly, the same work introduced a hybrid model combining attention and SSM layers. This hybrid model maintained faster inference times than standard transformers while achieving superior performance on all standard benchmarks tested, thus suggesting that hybrid architectures may be preferable.

Ultimately, the absolute scalability limits of pure SSM architectures remain an open question. While some  models today have successfully reached approximately 8 billion parameters \cite{waleffe2024empirical}, it is still unknown how far SSMs can be pushed before encountering diminishing returns. Meanwhile, a promising frontier has emerged in the form of hybrid architectures. By fusing traditional attention mechanisms with state-space components, some works have shown that these hybrid models can outperform both SSMs and transformers of similar size \cite{ren2024samba, lenz2025jamba}.

As the field progresses, however, control-theoretic techniques have yet to be used with SSMs. Given their structure, which naturally resembles nonlinear control systems, it remains an open and promising direction to investigate whether insights from control theory can be leveraged to analyze the limitations of SSMs and potentially improve their design and training.

\subsection{Safety and Alignment}

In recent years, the alignment of LLMs and SSMs with human intent has become a central challenge in deep learning. From a control-theoretic perspective, alignment can be formulated as a safety enforcement problem or as a trajectory tracking problem on a NN regarded as the plant.

Several efforts have adopted a feedback control perspective, even if this is not explicitly stated. One of the most popular techniques for alignment~\cite{ouyang2022training} uses human feedback to modify the underlying weights of a trained NN via reinforcement learning. This approach introduces a separate neural network, known as a reward model, which is trained to emulate human preferences. During the alignment phase, the language model acts as the plant, and the reward model provides the feedback signal used by a reinforcement learning algorithm to update the plant's internal parameters. This effectively closes the loop between model generation and human intent, steering the plant towards a preferred behavior.

Performing alignment during training is expensive since it delays the training process which is already quite expensive. A convenient alternative is \emph{test time alignment}, that is, algorithms that act as external controllers to regulate the NN behavior during inference without modifying the underlying weights. This is analogous to adding a controller to steer a given system to a reference. 

Control barrier functions (CBFs) have been used for test time alignment purposes. One framework \cite{miyaoka2025cbf} designs an add-on safety filter that intervenes in the token prediction process. The filter acts by filtering out undesirable words thus ensuring the generated text will not violate specified
safety constraints, such as toxicity thresholds. A different approach is taken in~\cite{11312450} by adding a one token prefix to every prompt. Control is exerted by steering this prefix so as to prevent toxic outputs. Toxicity is measured by a classifier that maps the state $z$ of an LLM into a numerical toxicity score for the text encoded by the tokens. The key result in~\cite{11312450} is that, under mild technical assumptions, a toxicity classifier defines a CBF for almost every choice of weights of an LLM with the added one token prefix as an input.

In~\cite{cheng2024linearly}, an optimal control approach is adopted to prevent toxicity. Text generation is framed as a trajectory in the space of tokens. The alignment is formulated as the problem of minimally changing this trajectory, subject to the token dynamics~\eqref{StateUpdate}, so as to steer the output of the activation functions away from regions that generate toxic outputs.

Often, preventing toxicity is but one of many objectives. Multi-objective decoding (MOD) \cite{mod_2024} addresses the challenge of satisfying multiple competing alignment goals simultaneously. By formulating the problem as maximizing a linear combination of a measure of goal satisfaction during the decoding step (the map $h$ in~\eqref{Composition} or~\eqref{TransformerEq}), this approach allows for the dynamic adjustment of these objectives at runtime, functioning like a variable-gain controller that can be tuned on the fly.

\subsection{Controllers for architectural and training design}

Feedback controllers have  been used to improve the behavior of certain architectures. A prime example is the PIDformer~\cite{nguyen2024pidformer} that inserts PID controllers in transformer blocks to create a new architecture. The objective of the PID controller is to prevent consensus, as discussed in~\emph{\nameref{Sec:LLMsTransformesConsensus}}, and improve robustness to input perturbations.

In addition to altering architectures, feedback controllers have been designed to improve the training of NNs. Research into PID-based optimizers has shown that applying PID control to adaptive gradient methods can dampen the oscillations seen in standard stochastic gradient descent or Adam method, leading to faster convergence \cite{wang2024pid}. Other approaches~\cite{wang2022incremental} treat learning rate scheduling as an online control problem, adjusting the optimizer's step size based on real-time feedback from the training loss via an incremental PID controller.  This line of work suggests that training algorithms can be seen as plants controlled by a feedback controller designed to drive the NN towards a minimum with desirable properties. Until now, only relatively simple control techniques have been used and more advanced control ideas are likely to have greater benefits.

\subsection{Contraction and implicit neural networks}

Contraction theory~\cite{FB-CTDS} has found multiple applications within control theory. Recently, its use has been expanded to also include the analysis of deep equilibrium networks~\cite{10.5555/3454287.3454350}. This is a class of NNs where state $z(k+1)$ of layer $k+1$ is computed from the state of layer $k$ using an implicit equation such as
\begin{equation}
\label{EqNet}
z(k+1)=\Sigma(W_1(k)z(k+1)+W_2(k)z(k)).
\end{equation}
One of the key challenges in this class of NNs is to ensure the weight matrices $W_1(k)$ and $W_2(k)$ are such that~\eqref{EqNet} admits a solution for $z(k+1)$. Since a solution can be interpreted as a fixed-point, the use of contraction ideas is natural and was employed in~\cite{10.5555/3540261.3541015} to establish existence of solutions as well as other properties. Recurrent versions of deep equilibrium networks have also been developed and, once again, contraction theory played a key role in their analysis~\cite{10179161}. The authors believe that contraction theory, and its generalizations such as $k$-contraction~\cite{WU2022110048}, can also play an important role in the analysis and design of other classes of NNs as well as their training algorithms.

\color{black}

\section{ACKNOWLEDGMENT}

The authors acknowledge Matteo Marchi, Rahal Nanayakkara, and Valen Yamamoto for valuable suggestions on how to improve this article. The 1st author was partially supported by grant PID2024-156578NB-I00 funded by MICIU/AEI/10.13039/501100011033/FEDER, EU. The work of the 2nd and 4th authors was partially supported by the NSF award 2502536. The work of the 3rd author is supported by the Natural Sciences and Engineering Research Council of Canada.

\section{Author Information}

\begin{IEEEbiography}{\'Alvaro {Rodr\'iguez Abella}}(arabella@comillas.edu) received the PhD degree in Mathematics from Complutense University of Madrid in Spain. He is currently an Assistant Professor of Applied Mathematics at Comillas Pontifical University. 
\end{IEEEbiography}

\begin{IEEEbiography}{Jo\~ao Pedro Silvestre}(joaosilvestre@ucla.edu) received the Masters degree in Electrical and Computer Engineering from Instituto Superior T{\'e}cnico in Lisbon, Portugal. He is currently pursuing his PhD on Electrical and Computer Engineering at UCLA. He is a Student Member of IEEE. 
\end{IEEEbiography}

\begin{IEEEbiography}{Bahman Gharesifard} (bahman.gharesifard@queensu.ca) received the PhD degree in Mathematics from Queen's University where he is currently a Professor of Mathematics. He is a Senior Member of IEEE. 
\end{IEEEbiography}

\begin{IEEEbiography}{Paulo Tabuada}{\,}(tabuada@ee.ucla.edu) received the PhD degree in Electrical and Computer Engineering from Instituto Superior T{\'e}cnico in Lisbon, Portugal. He is the Vijay K. Dhir Professor of Engineering at UCLA's Electrical and Computer Engineering Department. He is an IEEE Fellow. 
\end{IEEEbiography}

\bibliographystyle{ieeetr}
\bibliography{biblio}

\endarticle

\end{document}